\documentclass[11pt,a4paper]{article}
\usepackage[top=2cm,left=2cm,right=2cm]{geometry}

\usepackage[numbers,square]{natbib}

\usepackage{epsfig}
\usepackage{cancel}
\usepackage{caption}
\usepackage{feynmf} 
\usepackage{amssymb}
\usepackage{amsfonts}
\usepackage{epsf}
\usepackage{rotating}
\usepackage{graphicx}
\usepackage{amsmath}
\usepackage{fancyhdr}
\usepackage{subfigure}
\usepackage{graphics}
\usepackage{pstricks}
\usepackage{color}
\usepackage{frontespizio}
\usepackage{hyperref}
\hypersetup{
    colorlinks,
    citecolor=green,
    filecolor=black,
    linkcolor=blue,
    urlcolor=black
}
\usepackage{type1ec}
\usepackage[T1]{fontenc}
\usepackage{lettrine}
\usepackage{bbold}
\usepackage{calligra}
\usepackage{tikz}
\usepackage{subfigure}
\usepackage{mathrsfs}
\usepackage{curve2e}
\usepackage{setspace}
\usepackage{indentfirst}
\usepackage{emptypage}
\usepackage[babel]{csquotes} 
\usepackage[font=small,labelfont=bf,labelsep=quad]{caption} 
\usepackage{graphicx} 
\usepackage{listings} 
\usetikzlibrary{patterns}
\usepackage{relsize}
\usetikzlibrary{intersections,positioning}
\usetikzlibrary{decorations.pathmorphing,decorations.markings,arrows,positioning}
\usepackage{braket}
\usepackage{mathrsfs}
\usepackage{stackengine}
\usepackage{calc}
\newlength\shlength
\newcommand\xshlongvec[2][0]{\setlength\shlength{#1pt}%
  \stackengine{-5.6pt}{$#2$}{\smash{$\kern\shlength%
    \stackengine{7.55pt}{$\mathchar"017E$}%
      {\rule{\widthof{$#2$}}{.57pt}\kern.4pt}{O}{r}{F}{F}{L}\kern-\shlength$}}%
      {O}{c}{F}{T}{S}}

\newcommand{\sdfrac}[2]{\mbox{\small$\displaystyle\frac{#1}{#2}$}}

\IfFileExists{dsfont.sty}
{\usepackage{dsfont}
	\let\mathbb=\mathds
	\newcommand{\id}{\mathds{1}}}
{\typeout{Package dsfont.sty was not found, using alternative macros.}
	\let\mathds=\mathbb
	\newcommand{\id}{\mbox{1 \kern-.59em {\rm l}}}}

\usepackage{slashed}
\usepackage{units}
\usepackage{setspace}
\renewcommand{\Re}{\textrm{Re}}
\renewcommand{\Im}{\textrm{Im}}
\newcommand{\nn}{\nonumber}
\let\a=\alpha   \let\b=\beta   \let\g=\gamma   \let\d=\delta
         
    \let\k=\kappa  \let\l=\lambda  \let\m=\mu
\let\n=\nu      \let\x=\xi

\let\D=\Delta
\let\d=\delta

\newcommand{\figref}[1]{Fig.~\ref{#1}}			
\newcommand{\secref}[1]{Section~\ref{#1}}		
\newcommand{\appref}[1]{Appendix~\ref{#1}}		
\renewcommand{\a}{\alpha}

\newcommand{\G}{\Gamma}

\newcommand{\Ph}{\Phi}

\newcommand{\N}{\mathds{N}}

\def\nbox#1#2{\vcenter{\hrule \hbox{\vrule height#2in
			\kern#1in \vrule} \hrule}}
\def\sq{\,\raise.5pt\hbox{$\nbox{.09}{.09}$}\,}
\def\sqb{\,\raise.5pt\hbox{$\overline{\nbox{.09}{.09}}$}\,}

\newcommand{\bea}{\begin{eqnarray}}
\newcommand{\eea}{\end{eqnarray}}
\newcommand{\be}{\begin{equation}}
\newcommand{\ee}{\end{equation}}

\newcommand{\bes}{\begin{subequations}}
	\newcommand{\ees}{\end{subequations}}

\def\nn{\nonumber\\}

\numberwithin{equation}{section}

\usepackage{accents}

\begin{document}
\begin{center}
\vspace{1.5cm}
{\Large\bf On Some Hypergeometric Solutions }
\vspace{0.3cm}
{\Large  \bf of the Conformal Ward Identities of Scalar 4-point Functions in Momentum Space\\ }

\vspace{0.1cm}
\vspace{0.1cm}

\vspace{0.5 cm}
\vspace{0.2cm}
\vspace{0.3cm}

\vspace{2 cm}

{\bf  Claudio Corian\`o and Matteo Maria Maglio \\}

\vspace{0.5cm}

{\small \it Dipartimento di Matematica e Fisica "Ennio De Giorgi"\\
Universit\`a del Salento and INFN Lecce, \\
Via Arnesano, 73100 Lecce, Italy}

\vspace{0.5cm}

\end{center}

\begin{abstract}
We discuss specific hypergeometric solutions of the conformal Ward identities (CWI's) of scalar 4-point functions of primary fields in momentum space, in $d$ spacetime dimensions. We determine such solutions using various dual conformal ans\"atze (DCA's). We start from a  generic dual conformal correlator, and require it to be conformally covariant in coordinate space. The two requirements constrain such solutions to take a unique hypergeometric form. They describe correlators which are at the same time conformal and dual conformal in any dimension. These specific ans\"atze also show the existence of a link between 3- and 4-point functions of a CFT for such class of exact solutions, similarly to what found for planar ladder diagrams. We show that in $d=4$ only the box diagram and its melonic variants, in free field theory, satisfies such conditions, the remaining solutions being nonperturbative.   
We then turn to the analysis of some approximate high energy fixed angle solutions of the CWI's which also in this case take the form of generalized hypergeometric functions. We show that they describe the behaviour of the 4-point functions at large energy and momentum transfers, with a fixed $-t/s$. The equations, in this case, are solved by linear combinations of Lauricella functions of 3 variables and can be rewritten as generalized 4K integrals. 
In both cases the CWI's alone are sufficient to identify such solutions and their special connection with generalized hypergeometric systems of equations. 

\end{abstract}
\newpage

\section{Introduction} 
The study of conformal correlators of lower points, such as 2- and 3-point functions in $d =4$ and higher/lower spacetime dimensions, plays a special role in conformal field theory (CFT). In fact, they are almost completely determined by the symmetry of the theory, except for few constants which are specific of a given CFT.  One of the objectives of these investigations is to determine the correlation functions of a given theory without resorting to a Lagrangian realization. This allows to move beyond the standard perturbative approach, whenever this is possible, identifying interacting CFT's which do not necessarily have a corresponding free field theory realization.\\
One of the few reasons which motivate the study of such correlators in coordinate space is 
the possibility of imposing on them the conformal constraints in a simpler way compared to momentum space. 
This approach also plays a key role in the realization of the conformal  bootstrap, where higher point functions are computed starting from correlators of lower points, using an expansion in conformal partial waves \cite{Dolan:2000ut,Poland:2018epd}.\\
On the other hand, one of the advantages of the momentum space approach to the determination of CFT correlators, is that it allows to establish a link with the ordinary perturbative Feynman expansion. In particular, it allows to compare general results with explicit realizations of CFT's, where a large variety of methods are available. While the latter are directly connected with a specific Lagrangian realization, the analysis of the conformal Ward identities (CWI's) in momentum space, on the other hand, allows to investigate the operatorial content of a CFT in the most general way, whenever this is possible.
 As shown in the case of rather complex correlators such as the TTT and TJJ, where $T$ denotes the stress-energy tensor of a given CFT and $J$ a gauge current, by matching perturbative \cite{Coriano:2018bsy,Coriano:2018zdo,Coriano:2018bbe} and general CFT solutions \cite{2014JHEP...03..111B,Bzowski:2018fql,Bzowski:2015yxv,Bzowski:2017poo} - both in momentum space -  it is possible to rewrite the renormalized expressions of such correlators - in the most general CFT, at least for $d=4$ - in a very simple form, just in terms of scalar one-loop 2- and 3-point functions. \\
Most of such comparative studies performed in momentum space, which reconstruct the correlation function in a completely autonomous way, have dealt with scalar and tensor correlators \cite{Coriano:2018bsy,Coriano:2018bbe,2014JHEP...03..111B,Bzowski:2018fql,Bzowski:2015yxv,Coriano:2013jba,Bzowski:2015pba} only in $d=3,4,5$ dimensions, and limitedly to 3-point functions.\\ 
 In even spacetime dimensions, the study of such constraints in momentum space finds important applications in the context of the conformal anomaly action \cite{Giannotti:2008cv,Armillis:2009pq, Armillis:2010qk}, which has been investigated in the perturbative context in $d=4$.  \\
Beside the case of $d=4$, we also mention that in $d=3$ such correlators play an important role in the analysis of the gravitational perturbations and find wide applications in the investigation of nongaussianities \cite{Maldacena:2011nz}, in holographic cosmology \cite{Bzowski:2011ab,Coriano:2012hd} and also in condensed matter \cite{Chernodub:2017jcp}.
More recently, a general formalism for the extension of such analysis to a De Sitter background, in the context of inflationary cosmology,  has been formulated \cite{Arkani-Hamed:2018kmz,Benincasa:2018ssx,Arkani-Hamed:2018bjr,Arkani-Hamed:2017fdk}.  
The general analysis of such correlators provides a complementary approach with respect to those developed in the last two decades in the context of perturbative gauge theory amplitudes (see \cite{Henn:2014yza,Benincasa:2013faa} for an overview). The latter, in fact, rely on specific Lagrangian realizations and on supersymmetry. 
In $d=4$, one of the principal goals of this program, from our perspective, is the investigation of the structure of multipoint correlators containing stress energy tensors, in order to characterize the structure of the conformal anomaly action in a unique way.
\subsection{Towards 4-point functions}
In any attempt to move towards correlators of higher rank and spin, which can obviously parallel the significant developments obtained in coordinate space, it is necessary to investigate first the case of scalar amplitudes, as already done for 3-point functions. As well known from coordinate space, such 4-point correlators are not completely fixed by the CWI's, since these are easily solved modulo a generic function of the conformal invariant ratios. The ambitious goal of the conformal bootstrap program is to constrain such undetermined functions by using an expansion in conformal partial waves and the operator algebra of the corresponding OPE.\\
In some cases, however, it is possible to completely fix some correlation functions only by solving the corresponding CWI's, as we are going to show, by invoking an extra symmetry. This extra symmetry combines conformal invariance and dual conformal invariance at the same time. Also in this case it is important to work in full generality, only at the level of the CWI's, and we will derive the explicit forms of such solutions in momentum space. We will show that the CWI's reduce to an hypergeometric system of equations which can be solved as in the case of 3-point functions.\\

 \subsection{Dynamical symmetries in momentum space}
 Often, conformal properties of the perturbative expansion are found by the direct inspection of large classes of Feynman diagrams and indicate the presence of symmetries in the perturbative expansion of a given Lagrangian field theory. The simplest example is provided by ladder diagrams of 3- and 4-point functions in a scalar $\phi^3$ theory which, obviously, is not conformal invariant in $d=4$, due to a dimensionful coupling.\\
 However, as noticed long ago by Ussuykina, Davydychev and Broadhurst \cite{Davydychev:1992xr, Usyukina:1992jd,Usyukina:1993ch, Broadhurst:1993ru} a certain class of ladder diagrams of 3- and 4-point functions are related by certain redefinitions of some combinations of momenta in their explicit solutions. Such properties are not identified as generated by symmetries of the original Lagrangian, but provide - neverthless -  examples of other symmetries of the integrands of such diagrams, later denominated {\em dual conformal symmetries} (or DC). They play a role in specific field theory realizations, as in the planar limit of the $\mathcal{N}=4$ super Yang-Mills gauge theory, and are generically identified as being of dynamical origin. Obviously, such symmetries although characterized by a set of WI, are not symmetries of the action. \\  
 We establish a link, whenever such a link exists, between the results of \cite{Davydychev:1992xr, Usyukina:1992jd,Usyukina:1993ch, Broadhurst:1993ru} and the general CWI's of primary operators. Obvious differences exist between the class of conformal integrals identified in perturbation theory, which provide a realization of the DC symmetries in a Lagrangian realization, and the solutions of the CWI's of primary operators, which are rendered manifest by the type of ans\"atze that we choose. In particular, we will show that CFT' s where such symmetries are realized are essentially non-perturbative. For instance, ladder correlators do not share such symmetries, except for the box diagram and its {\em melonic} extensions.
 
 \subsection{Our work} 
In this work we are going to move to the analysis of 4-point functions in momentums space by investigating some scaling solutions of primary operators, showing that the hypergeometric character of the corresponding CWI's, already found in the case of 3-point functions, at least for such solutions, is preserved. Our analysis extends to  4-point functions  previous similar studies \cite{Coriano:2018bsy,Coriano:2018zdo,Coriano:2018bbe,2014JHEP...03..111B,Bzowski:2018fql,Bzowski:2015yxv,Coriano:2013jba}, formulated for scalar correlators. \\
 The solutions that we present, as we are going to elaborate, 
can be classified as being dual conformal (DC), as described in \cite{Drummond:2006rz,Drummond:2007aua,Drummond:2008vq}, {\em and} conformal in coordinate space (DCC) at the same time. They are constructed by requiring that they satisfy the first order differential conditions of dual conformal invariance, together with the second order ones coming from ordinary conformal symmetry. Both conditions are implemented and solved in momentum space.\\
The solutions that we identify can be written in two forms, either as generalized hypergeometrics, now functions of quartic ratios of momenta, or as integrals of 3 Bessel functions (3K integrals). It will be clear, from our approach, that a central part in our analysis is played by the hypergeometric system of partial differential equations (PDE's) which emerge from CWI's once we select a certain ans\"atz for a given correlator in momentum space.

\subsection{Dual conformal ans\"atz (DCA) and conformal invariance in coordinate space}
We use specific (dual conformal) ans\"atze (DCA's) to reduce the system of CWI's to Appell's hypergeometric functions, by introducing specific factorization of the expression of the correlators in terms of a scaling factor and of a remaining scale-invariant function of some conformal ratios. The various DCA's allow us to build such exact solutions in momentum space, link 3- and 4- point functions, exemplifying well known previous results  \cite{Davydychev:1992xr, Usyukina:1992jd,Usyukina:1993ch,Broadhurst:1993ru,Eden:1998hh,Eden:2000mv} on ladder diagrams in perturbation theory, as mentioned above. These have provided the first examples of dual conformal symmetry in the planar limit for scalar ladders.\\
We solve the equations in two cases, for equal scalings $(\Delta_i=\Delta, i=1,...4)$ of the primary operators and for two separate scalings $(\Delta_1=\Delta_2=\Delta_x, \Delta_3=\Delta_4=\Delta_y)$. 
The choice of the ans\"atz in momentum space implies that the solutions that we are looking for are dual conformal to begin with, and their Fourier transform to coordinate space is conformal as well. This last step is guaranteed if the ans\"atz satisfies the ordinary CWI's in coordinate space, which become second order PDE's in momentum space. 

We show that for the solution of CWI's in momentum space that we derive one can use the same formalism of the 3K integrals known for 3-point functions, though equivalent to their hypergeometric form.\\
By re-expressing the solutions generated by the ans\"atze as 3K integrals, the different ans\"atze are shown to determine a unique class of solutions, expressed just in terms of an overall constant and specific scaling dimensions. We will comment on the difference between such a result and those found in the computation of ladder diagrams in perturbation theories, where different dual conformal expressions - associated to specific one, two loop diagrams etc. - have, obviously, different analytic expressions.

\subsection{Approximate conformal solutions of primary operators and the Lauricella system}
Beside the search for exact solutions of the CWI's using the DCA in momentum space, in a second part of our  study we are also going to focus our attention on some approximate solutions of the same CWI's (for primary operators) using a specific kinematic approximation. Obviously, all our considerations apply to ordinary scattering amplitudes which are conformal in coordinate space, in particular to Feynman integrals of such type. We show that if we consider large $s$ and $t$ (Mandelstam) invariants in the correlators, with $-t/s$ fixed, suitable for a description of the same equations at fixed angle, the CWI's simplify.\\
The equations, in this approximation, are going to factorize the dependence on the external invariants $s, t$, from the remaining external mass invariants $p_i^2$. 
We show that the equations are fully compatible with asymptotic solutions which are logarithmic in  $-t/s$ in the Minkowski region, while the external mass invariants parameterize Lauricella functions, i.e. hypergeometric functions of 3 independent ratios. We show how such solutions and systems of equations can be equivalently described by the natural generalization of the 3K integrals to 4K. We conjecture that this pattern may extend to even higher point functions when the external mass invariants are separated from the remaining invariants scalar products of 2 different momenta. It seems clear that such factorized ans\"atze capture the essential behaviour of these correlators in some special kinematical limits, as it has been long known in the case of the Regge limit even at next-to-leading order in the gauge coupling, using conformal methods of t-channel unitarity \cite{Coriano:1995fj,Coriano:1996rj,Coriano:1994wk,Coriano:1995hx}. In all these cases the CFT constraints provide rather simple predictions compared to the explicit  NLO computations performed in QCD, with new partial waves appearing at NLO in the conformal reconstruction of the evolution (BFKL) kernel at the same order. 
\subsection{Notational remarks} 

We will be denoting with $x_i$ the coordinate dependence of a correlator. We will reserve the symbols $y_i$ to denote the dual coordinates in momentum space of the same correlator, while the (incoming) four-momenta will be denoted as $p_i$. The variables $x$ and $y$ (without any lower positional index $i$) will be used to denote ratios in momentum space expressed in terms of the invariants built out of the momenta $p_i$. Instead, the two invariant ratios in coordinate space, defined below, will be denoted as $u(x_i)$ and $v(x_i)$. The same invariant ratios in the dual conformal coordinates will be denoted as $u(y_i)$ and $v(y_i)$. The generators of the dual conformal symmetry will carry a $y_i$ dependence, such as $D(y_i), K^\kappa(y_i)$ for dilatation and special conformal transformations. Their versions in momentum space will be denoted as $D(p_i), K^\kappa(p_i)$, where in all these cases $y_i\equiv(y_1\ldots y_4)$, $x_i\equiv(x_1,\ldots x_4)$ and $p_i\equiv (p_1,\ldots p_4)$. As will be hopefully clear in the following, $K^\kappa(y)$, the special conformal generator in momentum space but in the dual conformal coordinates is a first order differential operator while $K^\kappa(p_i)$ is second order. 
\section{Three- and four-point functions from conformal invariance for correlators of primaries}
In order to clarify the new features of 4-point functions respect to correlators of lower points, we start our discussion by reviewing the case of such correlators in coordinate space. For 3-point functions we summarize the approach used in the analysis of primary scalar 3-point functions directly in momentum space, discussed in previous studies \cite{Coriano:2013jba}.
 We consider the simple case of a correlator of $n$ primary scalar fields $O_i(x_i)$, each of scaling dimension $\Delta_i$
\begin{equation}
\label{defop}
\Phi(x_1,x_2,\ldots,x_n)=\braket{O_1(x_1)O_2(x_2)\ldots O_n(x_n)}.
\end{equation}
Among these, 3- and 4-point functions (beside 2-point functions) in any CFT are significantly constrained in their general structure. Scalar 3-point functions of primary operators $\phi_i$ of scaling dimensions 
$\Delta_i$ $(i=1,2,3)$ are constrained to be of the form
\begin{equation}
\label{corr}
\langle \phi_1(x_1)\phi_2(x_2)\phi_3(x_3)\rangle =\frac{C_{123}}{ x_{12}^{\Delta_t - 2 \Delta_3}  x_{23}^{\Delta_t - 2 \Delta_1}x_{13}^{\Delta_t - 2 \Delta_2} },\qquad \Delta_t\equiv \sum_{i=1}^3 \Delta_i.
\end{equation}
  $C_{123}$ is a constant which specifies the CFT (the "CFT data"). For 4-point functions the constraints determine the structure of the correlator in a less effective way. In that case one identifies the two cross ratios 
\begin{equation}
\label{uv}
u(x_i)=\frac{x_{12}^2 x_{34}^2}{x_{13}^2 x_{24}^2} \qquad v(x_i)=\frac{x_{23}^2 x_{41}^2}{x_{13}^2 x_{24}^2}
\end{equation}
and the general solution can be written in the form 
\begin{equation}
\label{general}
\langle \phi_1(x_1)\phi_2(x_2)\phi_3(x_3)\phi_4(x_4)\rangle= h(u(x_i),v(x_i))\, \frac{1}{\left(x_{12}^2\right)^\frac{\Delta_1 + \Delta_2}{2}\left(x_{3 4}^2\right)^\frac{\Delta_3 + \Delta_4}{2}}
\end{equation}
where $h(u(x_i),v(x_i))$ remains unspecified. We are going to show that the equations may constrain $h(u(x_i),v(x_i))$ to take a specific form in momentum space, if we look for a specific ans\"atz.
 
For scalar correlators the special CWI's are given by first order differerential equations 
\begin{equation}
\label{SCWI0}
K^\kappa(x_i) \Phi(x_1,x_2,\ldots,x_n) =0
\end{equation}
with 
\begin{equation}
\label{transf1}
K^\kappa(x_i) \equiv \sum_{j=1}^{n} \left(2 \Delta_j x_j^\kappa- x_j^2\frac{\partial}{\partial x_j^\kappa}+ 2 x_j^\kappa x_j^\alpha \frac{\partial}
{\partial x_j^\alpha} \right)
\end{equation}
being the corresponding generator in coordinate space. The same operator, deprived of the scaling coefficients, will be denoted as $K_0^\kappa(x_i)$, i.e. 
\begin{equation}
K_0^\kappa(x_i) \equiv \sum_{j=1}^{n} \left(2 x_j^\kappa x_j^\alpha \frac{\partial}
{\partial x_j^\alpha} - x_j^2\frac{\partial}{\partial x_j^\kappa}\right).
\end{equation}
Conformal covariance and conformal invariance in coordinate space simply refer to the validity of \eqref{SCWI0} and of
\begin{equation}
 K_0^\kappa(x_i) \Phi(x_1,x_2,\ldots,x_n) =0
 \end{equation}
respectively. Denoting with 
\begin{equation}
 \Phi(p_1,\ldots p_{n-1},\bar{p}_n)=\langle O_1(p_1)\ldots O_n(\bar{p}_n)\rangle 
\end{equation}
and 
\begin{equation}
K^\kappa(p_i)\equiv\sum_{j=1}^{n-1}\left(2(\Delta_j- d)\frac{\partial}{\partial p_j^\kappa}+p_j^\kappa \frac{\partial^2}{\partial p_j^\alpha\partial p_j^\alpha} -2 p_j^\alpha\frac{\partial^2}{\partial p_j^\kappa \partial p_j^\alpha}\right)
\end{equation}
 the Fourier transform of \eqref{defop} and of \eqref{transf1} respectively, the form of the second order differential equations is given by
\begin{equation}
K^\kappa(p_i)\Phi(p_1,\ldots p_{n-1},\bar{p}_n)=0,\label{SCWI}
\end{equation}
where we have chosen $\bar{p}_n^\mu=-\sum_{i=1}^{n-1} p_i^\mu$ the n-th momentum, to be the linearly dependent one.
The action of the differential operators is realized on the shell of momentum conservation, where the 4-th momentum, conventionally, will be taken as dependent from the previous ones. Coming to the dilatation WI's, in our conventions, a scale-covariant 
function in coordinate space 
\begin{equation}
\phi(\lambda x_i)=\lambda^{-\Delta}\phi(x_i) 
\end{equation}
gives in momentum space 
\begin{equation}
\phi(\lambda p_1\ldots \lambda \bar{p}_n)=\lambda^{-\Delta'}\phi(p_1\ldots \bar{p}_n),
\end{equation}
with 
\begin{equation}
\Delta'\equiv \left(-\sum_{i=1}^n \Delta_i +(n-1) d\right)=-\Delta_t +(n-1) d.
\end{equation}
The corresponding equations are
\begin{equation}
\label{scale12}
D(x_i)
\Phi(x_1,\ldots x_n)=0
\end{equation}
with 
\begin{equation}
\label{scale11}
D(x_i)\equiv\sum_{i=1}^n\left( x_i^\alpha \frac{\partial}{\partial x_i^\alpha} +\Delta_i\right)
\end{equation}
for scale covariant correlators, in the case of scale invariance turn into 
\begin{equation}
D_0(x_i)
\Phi(x_1,\ldots x_n)=0
\end{equation}
with $D_0(x_i)$ given by
\begin{equation}
D_0(x_i)\equiv\sum_{i=1}^n\left( x_i^\alpha \frac{\partial}{\partial x_i^\alpha}\right). 
\end{equation}
In momentum space, the condition of scale covariance and invariance are respectively given by 
\begin{equation}
D(p_i) \Phi(p_1\ldots \bar{p}_n)=0
\end{equation}
with
\begin{equation}
D(p_i)\equiv\sum_{i=1}^{n-1}  p_i^\alpha \frac{\partial}{\partial p_i^\alpha} + \Delta'
\end{equation}
and 
\begin{equation}
D_0(p_i) \Phi(p_1\ldots \bar{p}_n)=0
\end{equation}
with
\begin{equation}
D_0(p_i)\equiv\sum_{i=1}^{n-1} p_i^\alpha \frac{\partial}{\partial p_i^\alpha} .
\end{equation}

Once we move to dual conformal coordinates in momentum space, denoted as $y_i$ below, it is important to keep clearly in mind the separation between actions of $K$ or $D$, such as those induced by their expressions in $x_i$ coordinates, from their second order in the $p_i$ variable. It is also common to refer to dual conformal symmetry to just an independent $SO(2,4)$ symmetry respect to the ordinary conformal symmetry of coordinate space (or of its Fourier image).

\subsection{Equations for 3-point functions and the hypergeometric solutions}
For 3-point functions the momentum dependence of the correlator is parameterized uniquely by $p_1^2, p_2^2$ and $p_3^3$, the three external invariant masses and we will denote with $p_i$ their magnitudes. The CWI's in momentum space, in this case, can be reduced to scalar equations by some manipulations, as discussed in \cite{Coriano:2018bbe,2014JHEP...03..111B,Coriano:2013jba}. 
 Introducing the operators 
\begin{subequations}
	\begin{align}
	\textup{K}_i &= \frac{\partial^2}{\partial p_i^2} + \frac{d+1-2\Delta_i}{p_i} \frac{\partial}{\partial p_i}  \qquad i=1,2,3    \\ 
	\textup{K}_{ij} &= \textup{K}_i - \textup{K}_j \,
	\end{align}\label{Koper}
\end{subequations}
$\Phi(p_1,p_2,p_3)$, in the scalar case, is constrained by two equations derived from the special conformal transformations
\begin{equation}
\textup{K}_{12}\Phi(p_1,p_2,p_3)=0 \qquad \qquad \textup{K}_{13}\Phi(p_1,p_2,p_3)=0 
\label{keq}
\end{equation}
and the dilatation WI
\begin{equation}
\label{scale}
\sum_{i=1}^3 p_i\frac{\partial}{\partial p_i} \Ph(p_1,p_2,p_3)=(\Delta_t-2 d) \Phi(p_1,p_2,p_3).
\end{equation}

Following the approach presented in \cite{Coriano:2013jba}, the ans\"atz for the solution can be taken of the form 
\begin{equation}
\label{ans}
\Phi(p_1,p_2,p_3)=p_3^{\,\Delta_t - 2 d} x^{a}y^{b} F(x,y)
\end{equation}
with $x=\frac{p_1^2}{p_3^2}$ and $y=\frac{p_2^2}{p_3^2}$. Here we are taking $p_3$ as "pivot" in the expansion, but we could have equivalently chosen as such any of the 3  momentum invariants $p_i^2$. $\Phi$ is required to be homogeneous of degree $\Delta_t-2 d$ under a scale transformation, according to \eqref{scale}, and in (\ref{ans}) this is taken into account by the factor $p_3^{\Delta_t - 2 d}$. 
In order to perform the reduction to the hypergeometric form of the equations, we need to set the (Fuchsian) indices
\begin{equation}
\label{cond1}
a=0\equiv a_0 \qquad \textrm{or} \qquad a=\Delta_1 -\frac{d}{2}\equiv a_1.
\end{equation}
In order to reduce the equation $\textup{K}_{13}\Phi=0$ to an hypergeometric system. From the equation $\textup{K}_{23}\Phi=0$ we obtain a similar condition for $b$, thereby fixing the two remaining (Fuchsian) indices
\begin{equation}
\label{cond2}
b=0\equiv b_0 \qquad \textrm{or} \qquad b=\Delta_2 -\frac{d}{2}\equiv b_1.
\end{equation}
 The complete equivalence of the CWI's \eqref{keq} with an hypergeometric system of equations is obtained by choosing such particular $(a,b)$ exponents in the non-scale invariant part of the ans\"atz. 
The four independent solutions of the CWI's then will all be characterized by the same 4 pairs of indices $(a_i,b_j)$ $(i,j=1,2)$.
Setting 
\begin{equation}
\alpha(a,b)= a + b + \frac{d}{2} -\frac{\Delta_1 +\Delta_2 -\Delta_3}{2} \qquad \beta (a,b)=a +  b + d -\frac{\Delta_1 +\Delta_2 +\Delta_3}{2} \qquad 
\label{alphas}
\end{equation}
the general solutions takes the form 
\begin{equation}
\Phi(p_1,p_2,p_3)=p_3^{\Delta-2 d}\, \sum_{a,b} c(a,b,\vec{\Delta_t})\,x^a y^b \,F_4(\alpha(a,b), \beta(a,b); \gamma(a), \gamma'(b); x, y)
\label{geneq} 
\end{equation}
where the sum runs over the four values $a_i, b_i$ $i=0,1$ with constants $c(a,b,\vec{\Delta_t})$ and $\vec{\Delta_t}=(\Delta_1,\Delta_2,\Delta_3)$. Defining 
\begin{align}
&\alpha\equiv \alpha(a_0,b_0)=\frac{d}{2}-\frac{\Delta_1 + \Delta_2 -\Delta_3}{2},\, && \beta\equiv \beta(b_0)=d-\frac{\Delta_1 + \Delta_2 +\Delta_3}{2},  \nn
&\gamma \equiv \gamma(a_0) =\frac{d}{2} +1 -\Delta_1,\, &&\gamma'\equiv \gamma(b_0) =\frac{d}{2} +1 -\Delta_2.
\end{align}
the 4 independent solutions can be re-expressed in terms of the parameters above as 
\begin{align}
\label{F4def}
S_1(\alpha, \beta; \gamma, \gamma'; x, y)\equiv F_4(\alpha, \beta; \gamma, \gamma'; x, y) = \sum_{n = 0}^{\infty}\sum_{m = 0}^{\infty} \frac{(\alpha)_{n+m} \, 
(\beta)_{n+m}}{(\gamma)_n \, (\gamma')_m} \frac{x^n}{n!} \frac{y^m}{m!} 
\end{align}
with the definition of the Pochhammer symbol $(\l)_{k}$ given by
\begin{equation}
(\l)_{k}=\frac{\G(\l+k)}{\G(\l)}=\l(\l+1)\dots(\l+k-1),\label{Pochh}
\end{equation}
and
\begin{align}
\label{solutions}
S_2(\alpha, \beta; \gamma, \gamma'; x, y) &= x^{1-\gamma} \, F_4(\alpha-\gamma+1, \beta-\gamma+1; 2-\gamma, \gamma'; x,y) \,, \nn
S_3(\alpha, \beta; \gamma, \gamma'; x, y) &= y^{1-\gamma'} \, F_4(\alpha-\gamma'+1,\beta-\gamma'+1;\gamma,2-\gamma' ; x,y) \,, \nn
S_4(\alpha, \beta; \gamma, \gamma'; x, y) &= x^{1-\gamma} \, y^{1-\gamma'} \, 
F_4(\alpha-\gamma-\gamma'+2,\beta-\gamma-\gamma'+2;2-\gamma,2-\gamma' ; x,y) \, . 
\end{align}
for which the solution can be written in the final form
\begin{equation}
\Phi(p_1,p_2,p_3)=p_3^{\Delta-2 d} \sum_{i=1}^4 \,c_i(\D_1,\D_2,\D_3)\,S_i (\alpha, \beta; \gamma, \gamma'; x, y)
\end{equation}
where $c_i$ are arbitrary coefficients which may depend on the scale dimensions $\D_i$ and on the spacetime dimension $d$. An equivalent version of the solution found above can be derived as in \cite{2014JHEP...03..111B}, where it is written in terms of $K$ Bessel functions as
\begin{equation}
\label{caz}
\Phi(p_1,p_2,p_3)=\,C_{123}\, p_1^{\D_1-\frac{d}{2}}p_2^{\D_2-\frac{d}{2}}p_3^{\D_3-\frac{d}{2}}\int_0^\infty dx\,x^{\frac{d}{2}-1}\,K_{\D_1-\frac{d}{2}}(p_1\,x)\,K_{\D_2-\frac{d}{2}}(p_2\,x)\,K_{\D_3-\frac{d}{2}}(p_3\,x)
\end{equation}
where $C_{123}$ is an undetermined constant. This formalism will be used later in the analysis of the solution of the 4-point function.

\subsection{Symmetrizations}
Notice that in the scalar case, for ordinary correlators, one is allowed to require its complete symmetry under the exchange of the 3 external momenta and scaling dimensions, as discussed in \cite{Coriano:2013jba}. This reduces the four
constants of integration to just one overall. The 4 independent solutions are then all of the form $x^a y^b F_4$, with
$a$ and $b$ fixed by (\ref{cond1}) and (\ref{cond2}). Such values of the $(a,b)$ exponents in the part of the ans\"atz which is not scale invariant, are determined by the condition that the $1/x$ and $1/y$ contributions vanish in the PDE's, turning the CWI's into a hypergeometric system of two equations, whose structure is symmetric under the exhange of $x$ and $y$.\\
For tensor correlators such as the $TJJ$ or the $TTT$ an extensive use of the properties of the the hypergeometric operators $\textup{K}_{ij}$ allows to build the complete solutions for the form factors which parameterize each of these correlators \cite{Coriano:2018bsy,Coriano:2018bbe}. Imposing the symmetry conditions is, in general, rather cumbersome, and one has to rely on one of the few relations known for the Appell function $F_4$, specifically the inversion formula 
\begin{align}
\label{transfF4}
F_4(\alpha, \beta; \gamma, \gamma'; x, y) =& \quad\frac{\Gamma(\gamma') \Gamma(\beta - \alpha)}{ \Gamma(\gamma' - \alpha) \Gamma(\beta)} (- y)^{- \alpha} \, F_4\left(\alpha, \alpha -\gamma' +1; \gamma, \alpha-\beta +1; \frac{x}{y}, \frac{1}{y}\right) \notag\\ 
&+  \frac{\Gamma(\gamma') \Gamma(\alpha - \beta)}{ \Gamma(\gamma' - \beta) \Gamma(\alpha)} (- y)^{- \beta} \, F_4\left(\beta -\gamma' +1, \beta ; \gamma, \beta-\alpha +1; \frac{x}{y}, \frac{1}{y}\right) \,
\end{align}
which allows to reverse the ratios respect to the momentum chosen as pivot. The symmetrization, obviously, allows to reduce the number of constants. 
\subsection{Extracting the physical solution}
In order to clarify this subtle point, we illustrate the possible methods that can be followed in order to identify the unique physical solution of the hypergeometric equations. \\
Notice, as already mentioned above, that the four solutions \eqref{F4def} and \eqref{solutions} define the 
basis into which {\em any} solution can be expanded. Such basis allows to generate 
by linear combination any function which is symmetric in the external momenta, under the condition that the constants 
$c_i(\D_1,\D_2,\D_3)$ are appropriately chosen. This is exactly what \eqref{transfF4} allows to achieve. In fact, by using \eqref{transfF4}, the general symmetric solution can be identified - modulo a single overall constant - in the form 
{\cite{Coriano:2013jba}
\begin{align}
&\braket{O(p_1)\,O(p_2)\,O(p_3)}=\big(p_3^2\big)^{-d+\frac{\Delta_t}{2}}\,C(\Delta_1,\Delta_2,\Delta_3,d)\notag\\
&\Bigg\{\Gamma\left(\Delta_1-\frac{d}{2}\right)\Gamma\left(\Delta_2-\frac{d}{2}\right)\Gamma\left(d-\frac{\Delta_1+\Delta_2+\Delta_3}{2}\right)\Gamma\left(d-\frac{\Delta_1+\Delta_2-\Delta_3}{2}\right)\notag\\
&\hspace{3cm}\times
\,F_4\,\left(\frac{d}{2}-\frac{\Delta_1+\Delta_2-\Delta_3}{2},d-\frac{\Delta_t}{2},\frac{d}{2}-\Delta_1+1,\frac{d}{2}-\Delta_2+1;x,y\right)\notag\\[2ex]
&\qquad+\,
\Gamma\left(\frac{d}{2}-\Delta_1\right)\Gamma\left(\Delta_2-\frac{d}{2}\right)\Gamma\left(\frac{\Delta_1-\Delta_2+\Delta_3}{2}\right)\Gamma\left(\frac{d}{2}+\frac{\Delta_1-\Delta_2-\Delta_3}{2}\right)\notag\\
&\hspace{3cm}\times x^{\Delta_1-\frac{d}{2}}\,F_4\,\left(\frac{\Delta_1-\Delta_2+\Delta_3}{2},\frac{d}{2}-\frac{\Delta_2+\Delta_3-\Delta_1}{2},\Delta_1-\frac{d}{2}+1,\frac{d}{2}-\Delta_2+1;x,y\right)\notag\\[2ex]
&\qquad+\,
\Gamma\left(\Delta_1-\frac{d}{2}\right)\Gamma\left(\frac{d}{2}-\Delta_2\right)\Gamma\left(\frac{-\Delta_1+\Delta_2+\Delta_3}{2}\right)\Gamma\left(\frac{d}{2}+\frac{-\Delta_1+\Delta_2-\Delta_3}{2}\right)\notag\\
&\hspace{3cm}\times\,y^{\Delta_2-\frac{d}{2}}\,F_4\,\left(\frac{\Delta_2-\Delta_1+\Delta_3}{2},\frac{d}{2}-\frac{\Delta_1-\Delta_2+\Delta_3}{2},\frac{d}{2}-\Delta_1+1,\Delta_2-\frac{d}{2}+1;x,y\right)\notag\\[2ex]
&\qquad+\,
\Gamma\left(\frac{d}{2}-\Delta_1\right)\Gamma\left(\frac{d}{2}-\Delta_2\right)\Gamma\left(\frac{\Delta_1+\Delta_2-\Delta_3}{2}\right)\Gamma\left(-\frac{d}{2}+\frac{\Delta_1+\Delta_2+\Delta_3}{2}\right)\notag\\
&\hspace{3cm}\times\,x^{\Delta_1-\frac{d}{2}}\,y^{\Delta_2-\frac{d}{2}}F_4\,\left(-\frac{d}{2}+\frac{\Delta_t}{2},\frac{\Delta_1+\Delta_2-\Delta_3}{2},\Delta_1-\frac{d}{2}+1,\Delta_2-\frac{d}{2}+1;x,y\right)\Bigg\}.\label{solfin}
\end{align}
One can verify that the symmetric solution above does not have any unphysical singularity in the physical region and it has the expected behaviour in the large momentum limit $p_3\gg p_1$, in agreement with the requirements discussed in \cite{Bzowski:2014qja}. In fact, one can check that the previous expression, in the limit $p_3\gg p_1$ (expressible also as $p_3^2,\,p_2^2\to\infty$ with $p_2^2/p_3^2\to1$ fixed), it behaves as
\begin{align}
\braket{O(p_1)\,O(p_2)\,O(p_3)}\propto f(d,\Delta_i)\,p_3^{\Delta_1+\Delta_2+\Delta_3-2d}\left( 1 +O\left(p_1/p_3\right)\right)  \hspace{2cm}\text{if}\ \ \Delta_1>\frac{d}{2}&
\end{align}
and
\begin{align}
\braket{O(p_1)\,O(p_2)\,O(p_3)}\propto g(d,\Delta_i)\,p_3^{\Delta_2+\Delta_3-\Delta_1-d}\,p_1^{2\Delta_1-d}\left(1 +O\left(p_1/p_3\right)\right)\hspace{2cm} \text{if}\ \ \Delta_1<\frac{d}{2}&,
\end{align}
with $f(d,\Delta_i)$ and $g(d,\Delta_i)$ depending only on the scaling and spacetime dimensions. Notice that this approach introduces the minimal set of independent 
solutions.  The result above in \eqref{solfin} is in complete agreement with the direct computation performed by Davydychev \cite{Davydychev:1992xr} of the generalized master integrals, obtained by a Fourier transform of \eqref{corr} and the use of the Mellin-Barnes method. \\

An alternative method consists in performing an explicit symmetrization of each of the four solutions and corresponding constants $c_j S_j$ $(j=1,2,3,4)$, obtained by permuting the $(p_i,\Delta_i)$ under the $\mathcal{S}_3$ permutation group. \\
 We remark that the method, in this case, introduces twenty-four functionally dependent contributions which, again, can be simplified by a repeated use of \eqref{transfF4}. In this case one discovers, after this simplification, that the symmetric solution so generated may manifest some unphysical singularities which disappear for a specific choice of the fundamental constants. A rather lengthy computation shows that the choice of such constants coincides with those presented in the solution \eqref{solfin}, originally given in \cite{Coriano:2013jba}, which involves the four basic solutions $S_j$ \eqref{F4def} and \eqref{solutions}.

An alternative approach is based on the formalism of the 3K integrals developed in \cite{2014JHEP...03..111B,Bzowski:2015yxv}, which for 3-point function is automatically symmetric. In this case the linear combination of the four solutions $S_i$ appearing in each 3K integral - as one can deduce from Eq. \eqref{3K} - has been checked to be free of unphysical singularities in the region of convergence.  \\
In the case of 4-point functions the only method which appears manageable is the explicit symmetrization of the fundamental solutions accompanied by the requirement that the 
symmetric expression is free of unphysical singularities in the pnysical domain. We will be illustrating this point in the following sections. 

\section{CWI's for scalar four-point functions}
From this subsection on we discuss an extension of the method summarized above to 4-point functions. We follow a similar strategy, by choosing a specific set of variables to characterize the ans\"atz for the solution of the corresponding PDE's. In the case of 3-point functions it is quite clear that the special CWI's are two equations and one can explicitly show that they remain independent after we introduce the ans\"atz \eqref{ans}. In the class of solutions that we look for, with a specific ans\"atz, two of the three constraining equations are independent, while a third equation is automatically satisfied.

In the case of the four point function the correlator depends on six invariants that we will normalize as
$p_i=|\sqrt{p_{i}\,{\hspace{-0.09cm}}^2}|$, $i=1,\dots,4$, representing the magnitudes of the momenta, and $s=|\sqrt{(p_1+p_2)^2}|$, $t=|\sqrt{(p_2+p_3)^2}|$ the two Mandelstam invariants, redefined by a square root. The CWI's are, in this case

\begin{equation}
\braket{O(p_1)\,O(p_2)\,O(p_3)\,O(\bar{p}_4)}=\Phi(p_1,p_2,p_3,p_4,s,t).\label{invariant}
\end{equation}
 
This correlation function, to be conformally invariant, has to verify the dilatation Ward Identity
\begin{equation}
\left[\sum_{i=1}^4\D_i-3d-\sum_{i = 1}^3p_i^{\mu}\frac{\partial}{\partial p_i^\mu}\right]\braket{O(p_1)\,O(p_2)\,O(p_3)\,O(\bar{p}_4)}=0
\end{equation}
and the special conformal Ward Identities
\begin{equation}
\sum_{i=1}^3\left[2(\D_i-d)\frac{\partial}{\partial p_{i\,\k}}-2p_i^\alpha\frac{\partial^2}{\partial p_i^\alpha\partial p_i^\kappa}+p_i^\kappa\frac{\partial^2}{\partial p_i^\alpha\partial p_{i\,\alpha}}\right]\braket{O(p_1)\,O(p_2)\,O(p_3)\,O(\bar{p}_4)}=0.
\end{equation}
One can split these equations in terms of the invariants of the four-point function written in \eqref{invariant}, by using the chain rules
\begin{align}
\frac{\partial}{\partial p_{1\,\mu}}&=\frac{p_1^\mu}{p_1}\frac{\partial}{\partial p_1}-\frac{\bar{p}_4^\mu}{p_4}\frac{\partial}{\partial p_4}+\frac{p_1^\mu+p_2^\mu}{s}\frac{\partial}{\partial s}\\
\frac{\partial}{\partial p_{2\,\mu}}&=\frac{p_2^\mu}{p_2}\frac{\partial}{\partial p_2}-\frac{\bar{p}_4^\mu}{p_4}\frac{\partial}{\partial p_4}+\frac{p_1^\mu+p_2^\mu}{s}\frac{\partial}{\partial s}+\frac{p_2^\mu+p_3^\mu}{t}\frac{\partial}{\partial t}
\end{align}
and similarly for the $p_3^\mu$ momentum, where $\bar{p}_4^\mu=-p_1^\mu-p_2^\mu-p_3^\mu$. From this prescription the dilatation WI becomes
\begin{align}
\bigg[(\D_t-3d)-\sum_{i=1}^4p_i\frac{\partial}{\partial p_i}-s\frac{\partial}{\partial s}-t\frac{\partial}{\partial t}\bigg]\Phi(p_1,p_2,p_3,p_4,s,t)=0,\label{Dilatation4}
\end{align}
with $\Delta_t=\sum_i \Delta_i$ is the total scaling,
and the special CWI's can be written as
\begin{equation}
\sum_{i=1}^3\ p_i^\kappa\, C_i=0\label{primary},
\end{equation}
where the coefficients $C_i$ are differential equations of the second order with respect to the six invariants previously defined. Being $p_1^\k,\ p_2^\k,\ p_3^\k$, in \eqref{primary} independent, 
we derive three scalar second order equations for each of the three $C_i$, which must vanish independently.\\
At this stage the procedure to simplify the corresponding equations is similar to the one described in \cite{Coriano:2018bsy,Coriano:2018bbe}. A lengthy computation allows to rewrite the equations in the form
\begin{align}
C_1&=\bigg\{\frac{\partial^2}{\partial p_1^2}+\frac{(d-2\D_1+1)}{p_1}\frac{\partial}{\partial p_1}-\frac{\partial^2}{\partial p_4^2}-\frac{(d-2\D_4+1)}{p_4}\frac{\partial}{\partial p_4}\notag\\[1.5ex]
&\qquad+\frac{1}{s}\frac{\partial}{\partial s}\left(p_1\frac{\partial}{\partial p_1}+p_2\frac{\partial}{\partial p_2}-p_3\frac{\partial}{\partial p_3}-p_4\frac{\partial}{\partial p_4}\right)+\frac{(\D_3+\D_4-\D_1-\D_2)}{s}\frac{\partial}{\partial s}\notag\\[1.5ex]
&\qquad+\frac{(p_2^2-p_3^2)}{st}\frac{\partial^2}{\partial s\partial t}\bigg\}\,\Phi(p_1,p_2,p_3,p_4,s,t)=0\label{C1}
\end{align}
for $C_1$ and
\begin{align}
C_2&=\bigg\{\frac{\partial^2}{\partial p_2^2}+\frac{(d-2\D_2+1)}{p_2}\frac{\partial}{\partial p_2}-\frac{\partial^2}{\partial p_4^2}-\frac{(d-2\D_4+1)}{p_4}\frac{\partial}{\partial p_4}\notag\\
&\qquad+\frac{1}{s}\frac{\partial}{\partial s}\left(p_1\frac{\partial}{\partial p_1}+p_2\frac{\partial}{\partial p_2}-p_3\frac{\partial}{\partial p_3}-p_4\frac{\partial}{\partial p_4}\right)+\frac{(\D_3+\D_4-\D_1-\D_2)}{s}\frac{\partial}{\partial s}\notag\\
&\qquad+\frac{1}{t}\frac{\partial}{\partial t}\left(p_2\frac{\partial}{\partial p_2}+p_3\frac{\partial}{\partial p_3}-p_1\frac{\partial}{\partial p_1}-p_4\frac{\partial}{\partial p_4}\right)+\frac{(\D_1+\D_4-\D_2-\D_3)}{t}\frac{\partial}{\partial t}\notag\\[1.2ex]
&\qquad+\frac{(p_2^2-p_4^2)}{st}\frac{\partial^2}{\partial s\partial t}\bigg\}\,\Phi(p_1,p_2,p_3,p_4,s,t)=0
\label{C2}
\end{align}

\begin{align}
C_3&=\bigg\{\frac{\partial^2}{\partial p_3^2}+\frac{(d-2\D_3+1)}{p_3}\frac{\partial}{\partial p_3}-\frac{\partial^2}{\partial p_4^2}-\frac{(d-2\D_4+1)}{p_4}\frac{\partial}{\partial p_4}\notag\\[1.5ex]
&\qquad+\frac{1}{t}\frac{\partial}{\partial t}\left(p_2\frac{\partial}{\partial p_2}+p_3\frac{\partial}{\partial p_3}-p_1\frac{\partial}{\partial p_1}-p_4\frac{\partial}{\partial p_4}\right)+\frac{(\D_1+\D_4-\D_2-\D_3)}{t}\frac{\partial}{\partial t}\notag\\[1.5ex]
&\qquad+\frac{(p_2^2-p_1^2)}{st}\frac{\partial^2}{\partial s\partial t}\bigg\}\,\Phi(p_1,p_2,p_3,p_4,s,t)=0\label{C3}
\end{align}
for $C_2$ and $C_3$, in agreement with \cite{Arkani-Hamed:2018kmz}. One of the two equations that we will be solving will be  
$C_{13}\equiv C_1- C_3=0$ and it is convenient to present it explicitly

\begin{align}
C_{13}&=\bigg\{\frac{\partial^2}{\partial p_1^2}+\frac{(d-2\D_1+1)}{p_1}\frac{\partial}{\partial p_1}-\frac{\partial^2}{\partial p_3^2}-\frac{(d-2\D_3+1)}{p_3}\frac{\partial}{\partial p_3}\notag\\
&\qquad+\frac{1}{s}\frac{\partial}{\partial s}\left(p_1\frac{\partial}{\partial p_1}+p_2\frac{\partial}{\partial p_2}-p_3\frac{\partial}{\partial p_3}-p_4\frac{\partial}{\partial p_4}\right)+\frac{(\D_3+\D_4-\D_1-\D_2)}{s}\frac{\partial}{\partial s}\notag\\
&\qquad+\frac{1}{t}\frac{\partial}{\partial t}\left(p_1\frac{\partial}{\partial p_1}+p_4\frac{\partial}{\partial p_4}-p_2\frac{\partial}{\partial p_2}-p_3\frac{\partial}{\partial p_3}\right)+\frac{(\D_2+\D_3-\D_1-\D_4)}{t}\frac{\partial}{\partial t}\notag\\[1.2ex]
&\qquad+\frac{(p_1^2-p_3^2)}{st}\frac{\partial^2}{\partial s\partial t}\bigg\}\,\Phi(p_1,p_2,p_3,p_4,s,t)=0.\label{Eq2}
\end{align}

\section{Dual Conformal/Conformal (DCC) examples}
 \label{dccsection}
Before moving to a discussion of the DCA's and the character of the solutions that we are going to identify, we turn to some specific examples of perturbative 4-point functions which are both conformal and dual conformal at the same time (DCC). We recall that a dual conformal integral \cite{Drummond:2006rz,Drummond:2007aua,Drummond:2008vq} is a Feynman integral which, once rewritten in terms of some dual coordinates, under the action of $K^\kappa$, is modified by factors which depend only on the coordinates of the external points. The reformulation of the ordinary momentum integral in terms of such dual coordinates can be immediately worked out by drawing the associated dual diagram. 
\begin{figure}[t]
	\centering
	\raisebox{-0.5\height}{\includegraphics[scale=0.3]{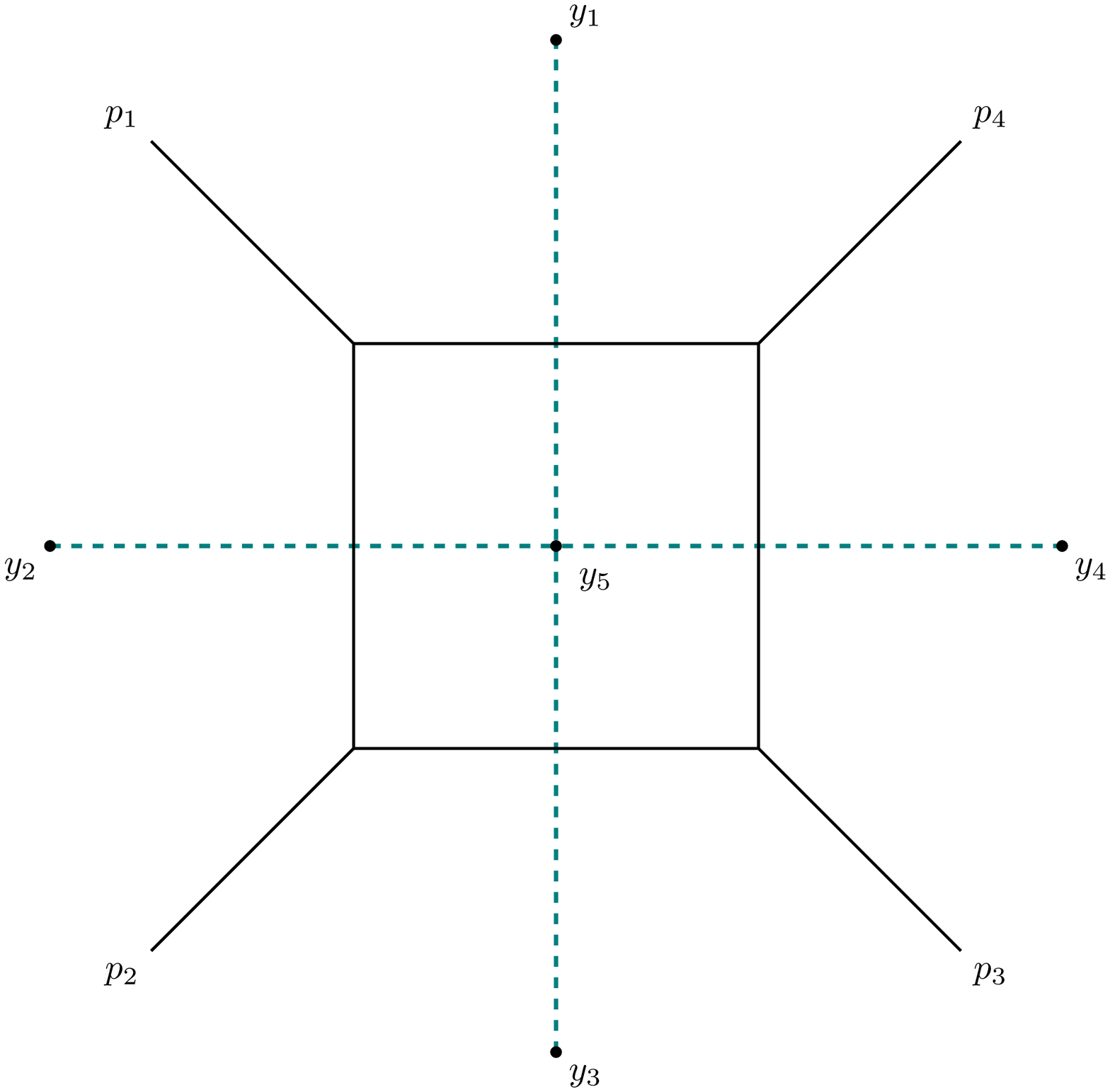}} \hspace{2cm}
	\raisebox{-0.5\height}{\includegraphics[scale=0.6]{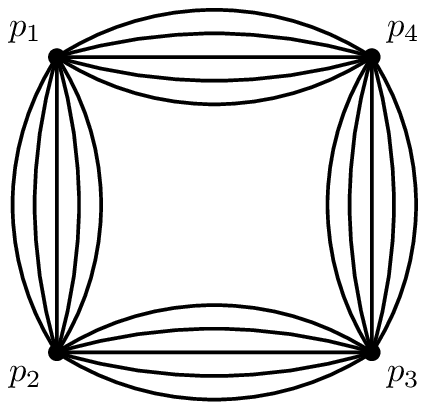}} 
	\caption{The box with its dual (left) and its higher scaling version (right). While the first is conformal for $d\ne2$ in ordinary conformal coordinates and for $d=4$ in dual coordinates, the right one is not conformal in coordinate and dual coordinate space at the same time. \label{Fig1}}
\end{figure}
We start from the ordinary box diagram (see Fig. \ref{Fig1})
\begin{equation}
\Phi_{Box}(p_1,p_2,p_3,p_4)=\int \frac{d^d k}{k^2 (k + p_1)^2 (k + p_1 + p_2)^2 ( k + p_1 + p_2 + p_3)^2}
\end{equation}
and apply the redefinition in terms of momentum variables $y_i$
\begin{equation}
k=y_{51}, \qquad p_1=y_{12}, \qquad p_2=y_{23},\qquad p_3=y_{34}
\end{equation}
with $y_{ij}=y_i-y_j$, thereby rewriting the integral in the form 
\begin{equation}
\label{box1}
\Phi_{Box}(y_1,y_2,y_3,y_4)=\int \frac{d^d y_5}{y_{15}^2 y_{25}^2 y_{35}^2 y_{45}^2}
\end{equation}
The action of $K^\kappa$ is realized in the form $\mathcal{I} \cdot \mathcal{T}\cdot \mathcal{I}$ (inversion, translation and inversion transformations) rather than as a differential action (by $K_0^\kappa$). We recall that under an inversion $(\mathcal{I})$ 
\begin{equation}
\mathcal{I} (d^d y_5)=d^d y_5 {(y_5^2)}^{-d}  \qquad \mathcal{I}(y^2_{ij})=\frac{y_{ij}^2}{y_i^2y_j^2}
\end{equation}
and in order to have an expression which is invariant under special conformal transformation, it is necessary to include a pre-factor in $\Phi_{Box}$, in the form  
\begin{equation}
 s^2 t^2 \Phi_{Box}(p_1,p_2,p_3,p_4)= y_{13}^2 y_{24}^2  \Phi_{Box}(y_1,y_2,y_3,y_4)
\end{equation}
 then its is easy to check that under the action of $\mathcal{I}$ the integrand 
 \begin{equation}
 \mathcal{I} \left(\frac{d^d y_5 y_{13}^2 y_{2 4}^2}{y_{15}^2 y_{25}^2 y_{35}^2 y_{45}^2}\right)= 
 \left(\frac{d^d y_5   (y_5^2)^{4-d}  y_{13}^2 y_{2 4}^2}{y_{15}^2 y_{25}^2 y_{35}^2 y_{45}^2}\right)
 \end{equation}
 becomes invariant under the action of the special conformal transformation if $d=4$. Obviously, the invariance under the complete action $\mathcal{I T I}$ is ensured. It is easily checked that the integrand is also scale invariant. It is then clear that the expression of the box diagram can only be of the form 
 \begin{equation}
 \label{ans1}
  \Phi_{Box}=\frac{1}{y_{13}^2 y_{24}^2}F\big(u(y_i),v(y_i)\big)
 \end{equation}
 with $u$ and $v$ given by 
 \begin{equation}
\label{uv2}
u(y_i)=\frac{y_{12}^2 y_{34}^2}{y_{13}^2 y_{24}^2} \qquad v(y_i)=\frac{y_{23}^2 y_{41}^2}{y_{13}^2 y_{24}^2}
\end{equation}
For future purposes it will be convenient to define
 \begin{equation}
\label{quartic}
x=\frac{p_1^2\,p_3^2}{s^2\,t^2},\qquad y=\frac{p_2^2\,p_4^2}{s^2\,t^2}
\end{equation}
 being the two invariant ratios $u(y_i),v(y_i)$, now expressed directly in terms of the original momentum invariants. 
 Notice that, by construction $u,v$ satisfy the first order equation in the $y$ variables
 \begin{equation}
\label{firstp}
\begin{split}
K_0^\kappa(y)\, u(y_i) &=\sum_{j=1}^{4} \left(- y_j^2\frac{\partial}{\partial y_j^\kappa}+ 2 y_j^\kappa y_j^\alpha \frac{\partial}
{\partial y_j^\alpha} \right)  u(y_i) =0\\
K_0^\kappa(y)\, v(y_i) &=\sum_{j=1}^{4} \left(- y_j^2\frac{\partial}{\partial y_j^\kappa}+ 2 y_j^\kappa y_j^\alpha \frac{\partial}
{\partial y_j^\alpha} \right)  v(y_i) =0
\end{split}
\end{equation}
while the action of $K_0^\kappa (p)$ on $x$ and $y$ will be nonzero.

 Notice that while the two forms of the $K_0^\kappa$ operator $K_0^\kappa(x_i)$ (coordinate) and $K_0^\kappa(p_i)$ (momenta) are one the Fourier transform of the other, $x$ and $y$ in \eqref{quartic} are not the Fourier images of $u(x_i)$ and $v(x_i)$. \\
 The box diagram is an example of a diagram which is dual conformal and conformal in $d=4$.  To show this point reconsider this diagram in coordinate space
 \begin{equation}
 \Phi_{Box}(x_i)=\frac{1}{x_{12}^2 x_{23}^2 x_{34}^2 x_{41}^2},
 \end{equation}
 that we can rewrite in the form 
\begin{align}
\label{dccc}
 \Phi_{Box}(x_i)&=\frac{1}{\left(x_{12}^2 x_{34}^2\right)^2}\left(\frac{x_{12}^2 x_{34}^2}{x_{23}^2 x_{41}^2}\right)\notag\\[1.2ex]
 &=\frac{1}{\left(x_{12}^2 x_{34}^2\right)^2}\left( \frac{u(x_i)}{v(x_i)}\right)
\end{align}
which is the conformally covariant correlator generated by 4 scalar primary fields $(\phi_i)$ in $d=4$ with $\Delta_i=2$. Denoting with $\chi$ an ordinary scalar field of scaling dimension 1, and setting $\phi_i=\chi^2$
we would have 
\begin{align}
\Phi_{Box}(x_i)&\equiv \langle \phi_1(x_1)\phi_2(x_2)\phi_3(x_3)\phi_4(x_4)\rangle = \langle \chi^2(x_1)\chi^2(x_2)\chi^2(x_3)\chi^2(x_4)\rangle \notag\\
&= \frac{1}{(x_{12}^{2})^\Delta (x_{34}^2)^\Delta}h\big(u(x_i),v(x_i)\big)
 \end{align}
with $\Delta=2$ and $h\big(u(x_i),v(x_i)\big)=u(x_i)/v(x_i)$. It is then obvious that the scalar box diagram satisfies the four constraints 
\begin{eqnarray}
& K^\kappa(x_i) \Phi(x_i)=0\qquad  D(x_i)\Phi_{Box}(x_i)=0\\
& K^\kappa(y_i) \Phi(y_i)=0\qquad  D(y_i)\Phi_{Box}(y_i)=0
\end{eqnarray}
in coordinates $x_i$ and dual (momentum) coordinates $y_i$ respectively as a system of first order PDE's. The system of equations can be all reported to momentum space in the form
\begin{align}
& K^\kappa(p_i) \Phi_{Box}(p_i)=0\qquad  D(p_i)\Phi_{Box}(p_i)=0\\
& K^\kappa(y_i) \Phi_{box}(y_i)=0\qquad  D(y_i)\Phi_{Box}(y_i)=0
\end{align}
as a system of second and first order constraints. 
We are going to discuss the solution of such constraints in detail, showing its unique hypergeometric structure.
\subsection{DCC solutions and the Feynman expansion: melonic contributions}
The case discussed above is a special one. In general, in fact, in perturbation theory,  it is possible to find solutions which are dual conformal or conformal, but not both, since some of the basic requirements are violated. \\
Consider the case of the perturbative melonic diagram shown in \figref{Fig1} where we have introduced a composite operator
\begin{equation}
\phi(x_i)=\chi^{n+m}(x_i)  \qquad  n,m\in  \N
\end{equation}
in $d$ dimensions with $n+m=N\in  \N$ fixed, which in free field theory generates the correlator 
\begin{equation}
\langle \phi(x_1)\phi(x_2)\phi(x_3)\phi(x_4)\rangle = \frac{1}{x_{12}^{2 a(n)} x_{23}^{2 b(m)}x_{34}^{2 a(n) }x_{41}^{2 b(m)}}
\end{equation}
with
\begin{equation}
a(n)=n \,\D,\quad b(m)=m \,\D,  \qquad \D= \frac{d-2}{2}
\end{equation}
which is conformally covariant since it can be re-expressed in the form 
\begin{equation}
\langle \phi(x_1)\phi(x_2)\phi(x_3)\phi(x_4)\rangle = \frac{1}{\left(x_{12}^2 x_{34}^{2} \right)^{a(n) + b(n)}}\left( 
\frac{u(x_i)}{v(x_i)}\right)^{b(n)}
\end{equation}
with the scaling dimension of $\phi$ given by $\left[\phi\right]= a(n) + b(m)$. In momentum space the corresponding integral is given by 
\begin{equation}
\int \frac{d^d k}{(k^2)^{\nu_1} ((k+p_1)^2)^{\nu_2} ((k+p_1+p_2)^2)^{\nu_3} ((k+ p_1 + p_2+p_3)^2)^{\nu_4}}
\end{equation}
with $\nu_1=\nu_3=d/2-a(n)$ and $\nu_2=\nu_4=d/2-b(m)$. Mapping this expression to dual coordinate, invariance of the integrand under special conformal transformations requires that 
\begin{equation}
m+ n=\frac{d}{d-2}
\end{equation}
which clearly shows that only $d=4$ allows to satisfy the dual conformal {\em and} conformal conditions, since $n+m$ has to be an integer. This brings us back to the ordinary box diagram. 

\subsection{DC symmetry and ladders}
We can slightly generalize the discussion presented above. It is convenient to introduce a more general notation, which can be used for the single, double etc. box diagrams, in order to set a distinction between correlators which are either dual conformal or conformal, or both. \\
The conformal behaviour of the box diagram in coordinate space $x_i$, for generic $d\ne2$ dimensions 
can be explicitly rewritten in the form
\begin{equation}
\Phi_{Box}(x_i)=\frac{1}{(x^2_{13})^{d-2}(x^2_{24})^{d-2}}\,\phi^{(1)}\left(u(x_i),v(x_i)\right),\quad d\ne2
\end{equation}
where $\phi^{(1)}\left(u(x_i),v(x_i)\right)$ is the undetermined function of the conformal ratios in coordinate space. 
$\phi^{(1)}$ can be easily identified from \eqref{dccc} in $d=4$ in perturbation theory. Its expression in dual (momentum space) coordinates can be rewritten as 
\begin{equation}
\Phi_{Box}(y_i)=\frac{1}{y^2_{13}\,y^2_{24}}\,\tilde{\phi}^{(1)}\left(u(y_i),v(y_i)\right),
\end{equation}
only in $d=4$. As elaborated above,  the box diagram can be both conformal and dual conformal invariant only in $d=4$. \\
Moving to the two-loop case, we consider the four-point ladder (planar) diagram (see Fig. \ref{dbbb}) and using the special conformal transformations, its expression takes the form
\begin{figure}[t]
	\centering
	\raisebox{-0.5\height}{\includegraphics[scale=0.4]{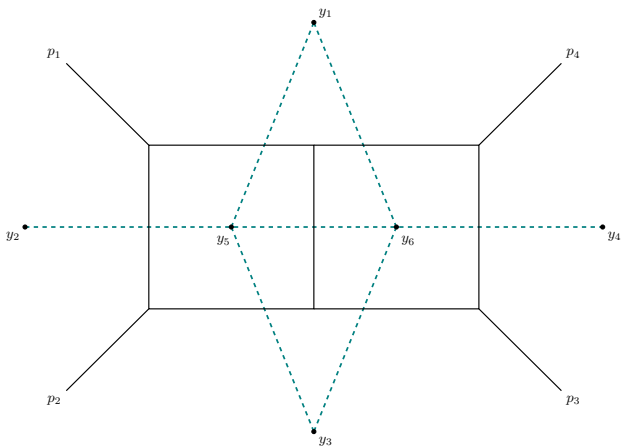}} 
	\caption{The two loop box digram with its dual. This diagram is not conformal in coordinate and dual coordinate space at the same time. }\label{dbbb}
\end{figure}
\begin{equation}
\Phi_{2-Box}(x_i)=\frac{1}{(x^2_{13})^{4}(x^2_{24})^{4}}\,\phi^{(2)}\left(u(x_i),v(x_i)\right),
\end{equation}
valid for $d=6$, where also in this case $\phi^{(2)}\left(u(x_i),v(x_i)\right)$ is another function of the conformal ratios in coordinate space, different from the one obtained in the one-loop case. Moving to momentum space and then to dual coordinates, we find the dual conformal expression of the double box in this space as
\begin{equation}
\Phi_{2-Box}(x_i)=\frac{1}{(y^2_{13})^{2}(y^2_{24})}\,\tilde{\phi}^{(2)}\left(u(y_i),v(y_i)\right),
\end{equation}
which holds for $d=4$.
It is obvious that the double box diagram can't be both conformal and dual conformal, at the same time and does not provide a perturbative realization of the solution previously found using the CWI's.\\ 
Using the same argument one can prove that the 4-point n-loop ladder diagram in coordinate space is conformal covariant   only in $d=6$, taking the form
\begin{equation}
\Phi_{n-Box}(x_i)=\frac{1}{(x^2_{13})^{4}(x^2_{24})^{4}}\,\phi^{(n)}\left(u(x_i),v(x_i)\right),
\end{equation}
valid for $n\ge 2$, where $\phi^{(n)}$ is a function of the conformal ratios. On the other hand, the same diagram in momentum space is dual conformal covariant  
\begin{equation}
\Phi_{n-Box}(x_i)=\frac{1}{(y^2_{13})^{n}\,y^2_{24}}\,\tilde{\phi}^{(n)}\left(u(y_i),v(y_i)\right), 
\end{equation}
for $n\ge 2$ and only for $d=4$. This shows that the class of solutions that we have identified are only realized at one-loop level.

\subsection{The triangle diagram}
The triangle diagram, on the other hand, is truly special, if we follow the same reasonings as above. Given its general expression
\begin{equation}
\label{j3}
J(\nu_1,\nu_2,\nu_3) = \int \frac{d^d l}{(2 \pi)^d} \frac{1}{(l^2)^{\nu_3} ((l+p_1)^2)^{\nu_2} ((l-p_2)^2)^{\nu_1}}\, ,
\end{equation}
with generic indices for the Feynman propagators $(\nu_1,\nu_2,\nu_3)$, it is easy to verify that the condition of dual 
conformal invariance 
\begin{equation}
\label{dci}
d=\nu_1+\nu_2 +\nu_3
\end{equation}
 allows to satisfy the DCC constraints in all dimensions. Such solutions are not obtained, in general, from free-field theories. We elaborate briefly on these points. A related discussion can be also found in \cite{Bzowski:2015pba}.\\
In fact, Eq. \eqref{j3} is the Fourier transform of a correlator of the form \eqref{corr}, for appropriate primary fields of scaling dimensions $\Delta_i$. Given some specific $\nu_i$, we can reverse-engineer three scalar primary fields of scalings $\Delta_i$ by the relations
\begin{equation}
\Delta_1 = d - \nu_2 - \nu_3 \,, \qquad
\Delta_2 = d - \nu_1 - \nu_3 \,, \qquad
\Delta_3 = d - \nu_1 - \nu_2, \,
\end{equation}
in such a way that \eqref{corr} is respected. Equivalently, 
\begin{equation}
\nu_1= \frac{1}{2} (d + \Delta_1 -\Delta_2-\Delta_3) \qquad
\nu_2 = \frac{1}{2}(d - \Delta_1 + \Delta_2 -\Delta_3)\,, \qquad
\nu_3 = \frac{1}{2}(d - \Delta_1 - \Delta_2 +\Delta_3).
\end{equation}

Using these relations, any conformal correlator of some scalar primaries of scaling $\Delta_i$'s, is 
bound to be of the form \eqref{corr}. The $\Delta_i's$  are trivially identified by the transform
\begin{align}
& \int \frac{d^d p_1}{(2\pi)^d} \frac{d^d p_2}{(2\pi)^d} \frac{d^d p_3}{(2\pi)^d} \, (2\pi)^d \delta^{(d)}(p_1 + p_2 + p_3) \, 
J(\nu_1,\nu_2,\nu_3) e^{- i p_1 \cdot x_1 - i p_2 \cdot x_2 - i p_3 \cdot x_3} \notag\\[1.5ex]
&\hspace{1cm} = \frac{1}{4^{\nu_1+\nu_2+\nu_3} \pi^{3 d/2}}  \frac{\Gamma(d/2 - \nu_1) \Gamma(d/2 - \nu_2) \Gamma(d/2 - \nu_3)}{\Gamma(\nu_1) 
\Gamma(\nu_2) \Gamma(\nu_3)} \Phi(x_1,x_2,x_3)  
\label{oner}
\end{align}
with 
\begin{equation}
 \Phi(x_1,x_2,x_3)\equiv \frac{1}{(x_{12}^2)^{d/2- \nu_3} (x_{23}^2)^{d/2- \nu_1} (x_{31}^2)^{d/2- \nu_2}}\,
\end{equation}
being the expression of a scalar conformal 3-point function.
Therefore, the conformal constraints in coordinated space on $\Phi(x_1,x_2,x_3)$ are automatically satisfied, providing no new information, while in momentum space they amount to some significant differential conditions
  \begin{equation}
  \label{cond12}
  \begin{split}
 K^\kappa(p_i) J(\nu_1,\nu_2,\nu_3) &=0 \\
 D(p_i) J(\nu_1,\nu_2,\nu_3)&=0
 \end{split}
 \end{equation}
 which need to be satisfied by the original integral $J$.\\
Eqs. \eqref{cond12} allow to obtain recursion relations among the class of master integrals associated to $J$  It can be also easily shown that the scale covariant condition, the second equation above, is equivalent to the integration by part rule used in the ordinary multiloop analysis of the master integrals \cite{Coriano:2013jba}.\\
We can follow a similar route with ordinary composite operators in free field theory, built out of scalar fields $\chi$ in $d$ dimensions, such as $\Phi=\chi^{2 n}$ with $\Delta_i=n(d-2)$. In this case the corresponding conformal 3-point function derived in free field theory is given by 
\begin{equation}
 \Phi(x_i)=\frac{1}{\left(x_{12}^2 x_{23}^2 x_{31}^2\right)^{n d'}}
 \end{equation}
$(d'=(d-2)/2)$ which generates the master integral $J$ with $\nu_i=d/2 -nd'$. If we require the dual conformal condition 
    \eqref{dci} to be valid, then this requires that $d= 6n/(3 n -1)$. For $d$ to be a physical dimension we require it to be an integer, and we are left only with the choice of $n=1$, which gives $d=3$. Therefore, the ordinary triangle diagram, if generated by a free CFT, is a DCC solution only in $d=3$.

\section{Factorized solutions of the CWI's  from DCA's}
Now we turn to discuss the solution of \eqref{C2} and of \eqref{Eq2}. As discussed above, we will consider possible solutions which are built around specific dual conformal ans\"atze, as illustrated in the previous sections. \\
The  equations involving $C_1$ and $C_3$ are both identically satisfied if the former equations \eqref{C2} and \eqref{Eq2} are. The number of independent equations, by using the ans\"atz that we are going to present below, will then reduce from 3 down to 2. We illustrate this procedure in some detail.

We choose  the ans\"atz
\begin{equation}
\Phi(p_i,s,t)=\big(s^2t^2\big)^{n_s}\,F(x,y)\label{ansatz}
\end{equation}
where $n_s$ is a coefficient (scaling factor of the ans\"atz) that we will fix below by the dilatation WI, and the variables $x$ and $y$ are defined by the quartic ratios
\begin{equation}
x=\frac{p_1^2\,p_3^2}{s^2\,t^2},\qquad y=\frac{p_2^2\,p_4^2}{s^2\,t^2}.
\end{equation}
 We will comment in a later section on the significance of such a choice and on the way to set up the invariants in momentum space in general. We will re-express the equations in terms of these new variables which will replace $s$ and $t$.\\
By inserting the ans\"atz \eqref{ansatz} into the dilatation Ward Identities, and turning to the new variables $x$ and $y$, after some manipulations we obtain from \eqref{Dilatation4} the condition
\begin{align}
\label{dil1}
&\bigg[(\D_t-3d)-\sum_{i=1}^4p_i\frac{\partial}{\partial p_i}-s\frac{\partial}{\partial s}-t\frac{\partial}{\partial t}\bigg]\big(s^2t^2\big)^{c}\,F(x,y),\notag\\
&=\big(s^2t^2\big)^{n_s}\big[(\D_t-3d)-4 n_s\big] \,F(x,y)=0
\end{align}
which determines $n_s=(\D_t-3d)/4$, giving
\begin{equation}
\Phi(p_i,s,t)=\big(s^2t^2\big)^{(\D_t-3d)/4}\,F(x,y).\label{ansatz2}
\end{equation}
We will be using this specific form of the solution in two of the three equations ($C_2$ and $C_{13}$). The functional form of $F(x,y)$ will then be furtherly constrained. 

\subsection{Determining the solutions in the case of primaries with equal scalings}
In order to determine the conditions on $F(x,y)$ from \eqref{C2} and \eqref{Eq2},  we re-express these two equations in terms of $x$ and $y$ using several identities. In particular we will use the relations 
\begin{equation}
\frac{\partial^2}{\partial s\partial t}F(x,y)=\frac{4}{st}\big[\big(
x\,\partial_x +y\partial_y\big)F+\big(x^2\partial_{xx}+2xy\,\partial_{xy}+y^2\partial_{yy}\big)F\big],
\label{onew}
\end{equation}
together with
\begin{align}
\left(p_1\frac{\partial}{\partial p_1}+p_2\frac{\partial}{\partial p_2}-p_3\frac{\partial}{\partial p_3}-p_4\frac{\partial}{\partial p_4}\right)F(x,y)&=\left(2x\,\partial_x+2y\,\partial_y -2x\,\partial_x-2y\,\partial_y\right)F(x,y)=0,\\[1.5ex]
\left(p_1\frac{\partial}{\partial p_1}+p_4\frac{\partial}{\partial p_4}-p_3\frac{\partial}{\partial p_3}-p_2\frac{\partial}{\partial p_2}\right)F(x,y)&=\left(2x\,\partial_x+2y\,\partial_y -2x\,\partial_x-2y\,\partial_y\right)F(x,y)=0.
\label{twow}
\end{align}
Both relations can be worked out after some lengthy computations using the relations presented in \appref{Appendix0}. \\
We start investigating the solutions of these equations by assuming, as a first example, that the scaling dimensions of all the fields $\phi_i$ are  equal $\D_1=\D_2=\D_3=\D_4=\D$.\\
 Using \eqref{onew} and \eqref{twow}, we write the first equation \eqref{C2} associated to $C_2$ in the new variable $x$ and $y$ as
\begin{align}
&C_2= 4\big(p_2^2-p_4^2\big)(s^2)^{n_s-1}(t^2)^{n_s-1}\notag\\
&\times\bigg[y(1-y)\partial_{yy} -2x\,y\,\partial_{xy}-x^2\partial_{xx}-(1-2n_s)x\,\partial_x+\left(1-\D+\frac{d}{2}-y(1-2n_s)\right)\,\partial_y-n_s^2\bigg] F(x,y)=0\label{first}
\end{align}
and the second one \eqref{Eq2} associated to $C_{13}$ as
\begin{align}
 &4\big(p_1^2-p_3^2\big)(s^2)^{n_s-1}(t^2)^{n_s-1}\notag\\
&\times\bigg[x(1-x)\partial_{xx} -2x\,y\,\partial_{xy}-y^2\partial_{yy}-(1-2n_s)y\,\partial_y+\left(1-\D+\frac{d}{2}-x(1-2n_s)\right)\,\partial_x-n_s^2\bigg] F(x,y)=0\label{second}
\end{align}
where we recall that $n_s$ is the scaling under dilatations, now given by
\begin{equation}
\label{scaled}
n_s=\D-\frac{3d}{4}
\end{equation}
since $\D_t=4\, \D$. \\
By inspection, one easily verifies that \eqref{first} and \eqref{second} define a hypergeometric system of two equations whose solutions can be expressed as linear combinations of 4 Appell functions of two variables $F_4$, as in the case of 3-point functions discussed before. The general solution of such system is expressed as
\begin{align}
\Phi(p_i,s,t) &=\big(s^2t^2\big)^{(\D_t-3d)/4}\, F(x,y) \notag\\
F(x,y)&= \sum_{a,b} c(a,b,\vec\Delta_t) x^a y^b F_4\left(\a(a,b),\b(a,b),\g(a),\g'(b);x, y\right),\label{solution}
\end{align}
with $\vec\Delta_t=\Delta(1,1,1,1)$ for being in the equal scaling case.  Notice that the solution is similar to that of the 3-point functions given by \eqref{geneq}, discussed before.\\
The general solution \eqref{solution} has been written as a linear superposition of these with independent constants $c(a,b)$, labelled by the exponents $a,b$ 
\begin{align}
&a=0,\,\D-\frac{d}{2},&&b=0,\,\D-\frac{d}{2},\label{FuchsianPoint}
\end{align}
which fix the dependence of the $F_4$  
\begin{align}
\label{s2}
&\a(a,b)=\frac{3}{4}d-\D+a+b,&&\b(a,b)=\frac{3}{4}d-\D+a+b,\notag\\
&\g(a)=\frac{d}{2}-\D+1+2a,&&\g'(b)=\frac{d}{2}-\D+1+2b.
\end{align}

We are now going to show that the third CWI corresponding to $C_1$ is identically satisfied by choosing the solution identified in \eqref{solution}. For this purpose we re-express the $C_1$ equation \eqref{C1} in terms of the $x$ and $y$ invariant ratios in the form 
\begin{equation}
\resizebox{1\hsize}{!}{$
	\begin{aligned}
C_1&=4(p_2^2-p_3^2)(s^2)^{n_s-1}(t^2)^{n_s-1}\left[x^2 \partial_{xx} +2x\,y\,\partial_{xy}+y^2\,\partial_{yy}+(1-2n)x\,\partial_x+(1-2n)y\,\partial_y+n^2\right]\,F(x,y) \\[1.3ex]
&+\frac{4\,(s^2)^{n_s}(t^2)^{n_s}}{p_1^2}\bigg[\,x^2\,\partial_{xx}-\frac{p_1^2}{p_4^2}\,y^2\,\partial_{yy}+\frac{(d-2\D+2)}{2}\,x\,\partial_x -\frac{(d-2\D+2)}{2}\frac{p_1^2}{p_4^2}\,y\,\partial_y \bigg]F(x,y) = 0.\label{C1mod}
\end{aligned}$}
\end{equation}
One can observe that the first line of the previous expression is actually a linear combination of the \eqref{first} and \eqref{second}. After some lengthy algebra we can rewrite the equation coming from $C_1$ in the form
\begin{equation}
(p_2^2+p_3^2)\bigg[2x\,\partial_{xx}-2y\,\partial_{yy}+(d-2\D+2)\partial_x-(d-2\D+2)\partial_y\bigg]F(x,y)=0.\label{C1new}
\end{equation}
In order to verify that the equation above is identically satisfied,  we use the following identities for the Appell hypergeometric function 
\begin{align}
\partial_x\,F_4(a,b,c_1,c_2;x,y)&=\frac{a\,b}{c_1}\,F_4(a+1,b+1,c_1+1,c_2;x,y)\\
\partial_y\,F_4(a,b,c_1,c_2;x,y)&=\frac{a\,b}{c_2}\,F_4(a+1,b+1,c_1,c_2+1;x,y)
\end{align}
\begin{equation}
x\,\partial_x\, F_4(a,b,c_1,c_2;x,y)= (c_1-1)\big[F_4(a,b,c_1-1,c_2;x,y)-F_4(a,b,c_1,c_2;x,y)\big].
\end{equation} 
We can use these relations to derive the further relation 
\begin{align}
x\,\partial_{xx}\, F_4(a,b,c_1,c_2;x,y)&=(c_1-1) \partial_x\,F_4(a,b,c_1-1,c_2;x,y)-c_1\,\partial_x\,F_4(a,b,c_1,c_2;x,y)\notag\\
&=a\,b\big[F_4(a+1,b+1,c_1,c_2;x,y)-F_4(a+1,b+1,c_1+1,c_2;x,y)\big]
\end{align}
with an analogous expression obtained for the $y$ variable. Considering the general expression of $F(x,y)$ previously obtained in \eqref{solution}, as $F(x,y)=x^a\,y^b\,F_4\left(\a(a,b),\b(a,b),\g(a),\g'(b);x, y\right)$ into \eqref{C1new} one indeed verifies that the equation
\begin{equation}
\resizebox{1\hsize}{!}{$
\begin{aligned}
0&=\bigg[2x\,\partial_{xx}-2y\,\partial_{yy}+(d-2\D+2)\partial_x-(d-2\D+2)\partial_y\bigg]x^a\,y^b\,F_4\left(\a(a,b),\b(a,b),\g(a),\g'(b);x, y\right)\\
&=x^a\,y^b\bigg[2x\,\partial_{xx}-2y\,\partial_{yy}+(d-2\D+2+2a)\partial_x-(d-2\D+2+2b)\partial_y\bigg]\,F_4\left(\a(a,b),\b(a,b),\g(a),\g'(b);x, y\right)
\end{aligned}$}
\end{equation}
is satisfied, if we choose $\a(a,b)$, $\b(a,b)$, $\g(a)$ and $\g'(b)$ as identified from \eqref{s2}.

Therefore one indeed verifies that equation $C_1$ vanishes on the chosen ans\"atz.

\subsection{Two independent operatorial scalings }
The solution obtained above in the equal scaling case can be extended to the more general case 
\begin{equation}
\D_1=\D_3=\D_x,\qquad \D_2=\D_4=\D_y.
\end{equation}
 In this case the CWI's  give the system of equations
\begin{equation}
\resizebox{1\hsize}{!}{$
\left\{\begin{aligned}
&\bigg[y(1-y)\partial_{yy} -2x\,y\,\partial_{xy}-x^2\partial_{xx}-(1-2n_s)x\,\partial_x+\left(1-\D_y+\frac{d}{2}-y(1-2n_s)\right)\,\partial_y-n_s^2\bigg] F(x,y)=0\\[2ex]
&\bigg[x(1-x)\partial_{xx} -2x\,y\,\partial_{xy}-y^2\partial_{yy}-(1-2n_s)y\,\partial_y+\left(1-\D_x+\frac{d}{2}-x(1-2n_s)\right)\,\partial_x-n_s^2\bigg] F(x,y)=0
\end{aligned}\right.$}
\end{equation}
where now $n_s$ is defined as 
\begin{equation}
n_s=\frac{\D_x}{2}+\frac{\D_y}{2}-\frac{3}{4}d.
\end{equation}
whose solutions are expressed as
\begin{equation}
\label{ress}
\Phi(p_i,s,t)=\big(s^2t^2\big)^{(\D_t-3d)/4}\,\sum_{a,b} c(a,b,\vec\Delta_t) x^a y^bF_4\left(\a(a,b),\b(a,b),\g(a),\g'(b);x,y\right)
\end{equation}
with $\vec\Delta_t=(\Delta_x,\Delta_y,\Delta_x,\Delta_y)$, $\Delta_t=2 \Delta_x + 2 \Delta_y$ and the Fuchsian points are fixed by the conditions
\begin{align}
&a=0,\,\D_x-\frac{d}{2}&&b=0,\,\D_y-\frac{d}{2}\notag\\
&\a(a,b)=\frac{3}{4}d-\frac{\D_x}{2}-\frac{\D_y}{2}+a+b,&&\b(a,b)=\frac{3}{4}d-\frac{\D_x}{2}-\frac{\D_y}{2}+a+b,\notag\\
&\g(a)=\frac{d}{2}-\D_x+1+2a,&&\g'(b)=\frac{d}{2}-\D_y+1+2b.
\end{align}
We pause for a moment to discuss the domain of convergence of such solutions. Such domain, for $F_4$, is bounded by the relation 
\begin{equation}
\sqrt{x}+\sqrt{y}< 1, 
\end{equation}
which is satisfied in a significant kinematic region, and in particular at large energy and momentum transfers. Notice that the analytic continuation of \eqref{ress} in the physical region can be simply obtained by sending $t^2\to -t^2$ (with $t^2<0$) and leaving all the other invariants untouched. In this case we get
\begin{equation}
 \sqrt{p_1^2 p_3^2}  +\sqrt{p_2^2 p_4^2} < \sqrt{- s^2 t^2}.
\end{equation}
At large energy and momentum transfers the correlator exhibits a power-like behaviour of the form 

\begin{equation}
\Phi(p_i,s,t)\sim \frac{1}{(- s^2 t^2)^{(3 d - \Delta_t)/4}}. 
\end{equation}
Given the connection between the function $F_4$ and the 3K integrals, we will reformulate this solution in terms of such integrals. They play a key role in the solution of the CWI's for tensor correlators, as discussed in \cite{2014JHEP...03..111B} for 3-point functions.

\subsection{DCC solutions as 3K integrals}
The link between 3- and 4-point functions outlined in the previous section allows to re-express the solutions in terms of a class of parametric integrals of 3 Bessel functions, as done in the case of the  scalar and tensor correlators  \cite{2014JHEP...03..111B}, with the due modifications.
We consider the case of the solutions characterized by $\D_1=\D_2=\D_3=\D_4=\D$ or $\D_1=\D_3=\D_x$ and\  $\D_2=\D_4=\D_y$. We will show that the solution can be written in terms of triple-K integrals which are connected to the Appell function $F_4$ by the relation 

\begin{align}
& \int_0^\infty d x \: x^{\alpha - 1} K_\lambda(a x) K_\mu(b x) K_\nu(c x) =\frac{2^{\alpha - 4}}{c^\alpha} \bigg[ B(\lambda, \mu) + B(\lambda, -\mu) + B(-\lambda, \mu) + B(-\lambda, -\mu) \bigg], \label{3K}
\end{align}
where
\begin{align}
B(\lambda, \mu) & = \left( \frac{a}{c} \right)^\lambda \left( \frac{b}{c} \right)^\mu \Gamma \left( \frac{\alpha + \lambda + \mu - \nu}{2} \right) \Gamma \left( \frac{\alpha + \lambda + \mu + \nu}{2} \right) \Gamma(-\lambda) \Gamma(-\mu) \times \notag\\
& \qquad \times F_4 \left( \frac{\alpha + \lambda + \mu - \nu}{2}, \frac{\alpha + \lambda + \mu + \nu}{2}; \lambda + 1, \mu + 1; \frac{a^2}{c^2}, \frac{b^2}{c^2} \right), \label{3Kplus}
\end{align}
valid for
\begin{equation}
\Re\, \alpha > | \Re\, \lambda | + | \Re \,\mu | + | \Re\,\nu |, \qquad \Re\,(a + b + c) > 0 \nn
\end{equation}
and the Bessel functions $K_\nu$ satisfy the equations 
\begin{align}
\frac{\partial}{\partial p}\big[p^\b\,K_\b(p\,x)\big]&=-x\,p^\b\,K_{\b-1}(p x)\nn
K_{\b+1}(x)&=K_{\b-1}(x)+\frac{2\b}{x}K_{\b}(x). \label{der}
\end{align}
In particular  the solution can be written as
\begin{equation}
I_{\a\{\b_1,\b_2,\b_3\}}(p_1\,p_3; p_2\,p_4;s\,t)=\int_0^\infty\,dx\,x^\a\,(p_1\,p_3)^{\b_1}\,(p_2\,p_4)^{\b_2}\,(s\,t)^{\b_3}\,K_{\b_1}(p_1\,p_3\,x)\,K_{\b_2}(p_2\,p_4\,x)\,K_{\b_3}(s\,t\,x).\label{trekappa}
\end{equation}
 Using  \eqref{der} one can derive several relations, 
such as 
\begin{align}
\frac{\partial^2}{\partial p_1^2}I_{\a\{\b_1,\b_2,\b_3\}}&=-\,p_3^2\,I_{\a+1\{\b_1-1,\b_2,\b_3\}}+p_1^2\,p_3^4\,\,I_{\a+2\{\b_1-2,\b_2,\b_3\}}
\end{align}
which generate identities such as 
\begin{align}
p_1^2\,p_3^2\,I_{\a+2\{\b_1-2,\b_2,\b_3\}}&=I_{\a+2\{\b_1,\b_2,\b_3\}}-2(\b_1-1)\,I_{\a+1\{\b_1-1,\b_2,\b_3\}}.
\end{align}
We refer to \appref{AppendixA} for more details and a complete list of identities for such integrals. Using these relations, the dilatation Ward identities \eqref{Dilatation4} take the form
\begin{equation}
(\D_t-3d) I_{\a\{\b_1,\b_2,\b_3\}}+2p_1^2p_3^2\ I_{\a+1\{\b_1-1,\b_2,\b_3\}}+2p_2^2p_4^2\ I_{\a+1\{\b_1,\b_2-1,\b_3\}}+2s^2t^2\ I_{\a+1\{\b_1,\b_2,\b_3-1\}}=0
\end{equation}
where the arguments of the $I_{\a\{\b_1\b_2\b_3\}}$ function, written explicitly in \eqref{trekappa}, have been omitted for the sake of simplicity. The $I$ integrals satisfy the differential equations
\begin{align}
\frac{1}{s}\frac{\partial}{\partial s}\left(p_1\frac{\partial}{\partial p_1}+p_2\frac{\partial}{\partial p_2}-p_3\frac{\partial}{\partial p_3}-p_4\frac{\partial}{\partial p_4}\right)I_{\a\{\b_1,\b_2,\b_3\}}&=0\\
\frac{1}{t}\frac{\partial}{\partial t}\left(p_1\frac{\partial}{\partial p_1}+p_4\frac{\partial}{\partial p_4}-p_2\frac{\partial}{\partial p_2}-p_3\frac{\partial}{\partial p_3}\right)I_{\a\{\b_1,\b_2,\b_3\}}&=0
\end{align} 
which can be checked using the relations given in the same appendix, and we finally find
\begin{equation}
(\D_t-3d+2\a+2-2\b_t) I_{\a\{\b_1,\b_2,\b_3\}}=0
\end{equation}
where $\b_t=\b_1+\b_2+\b_3$. In order to satisfy this equation the $\a$ parameter has to be equal to a particular value given by 
\begin{equation}
\tilde{\a}\equiv \frac{3}{2}d+\b_t-1-\frac{\D_t}{2}.
\end{equation}
 In the particular case $\D_i=\D$ the special conformal Ward identities are given by
\begin{equation}
\left\{\begin{aligned}&\\[-1.2ex]
&\bigg [\frac{\partial^2}{\partial p_1^2}+\frac{(d-2\D+1)}{p_1}\frac{\partial}{\partial p_1}-\frac{\partial^2}{\partial p_3^2}-\frac{(d-2\D+1)}{p_3}\frac{\partial}{\partial p_3}+\frac{(p_1^2-p_3^2)}{st}\frac{\partial^2}{\partial s\partial t}\bigg ]\,I_{\tilde\a\{\b_1,\b_2,\b_3\}}=0\\[1ex]
&\bigg[\frac{\partial^2}{\partial p_2^2}+\frac{(d-2\D+1)}{p_2}\frac{\partial}{\partial p_2}-\frac{\partial^2}{\partial p_4^2}-\frac{(d-2\D+1)}{p_4}\frac{\partial}{\partial p_4}+\frac{(p_2^2-p_4^2)}{st}\frac{\partial^2}{\partial s\partial t}\bigg ]\,I_{\tilde\a\{\b_1,\b_2,\b_3\}}=0\\[1ex]
&\bigg[\frac{\partial^2}{\partial p_3^2}+\frac{(d-2\D+1)}{p_3}\frac{\partial}{\partial p_3}-\frac{\partial^2}{\partial p_4^2}-\frac{(d-2\D+1)}{p_4}\frac{\partial}{\partial p_4}+\frac{(p_2^2-p_1^2)}{st}\frac{\partial^2}{\partial s\partial t}\bigg ]\,I_{\tilde\a\{\b_1,\b_2,\b_3\}}=0\\[-0.7ex]
\label{neweq}
\end{aligned}\right.
\end{equation}
and using the properties of Bessel functions they can be rewritten in a simpler form. The first equation, for instance, can be written as
\begin{equation}
\label{oone}
(p_1^2-p_3^2)\bigg( (d-2\D+2\b_1)\,I_{\tilde\a+1\{\b_1-1,\b_2,\b_3\}}-2\b_3\,I_{\tilde\a+1\{\b_1,\b_2,\b_3-1\}} \bigg)=0,
\end{equation}
which is identically satisfied if the conditions 
\begin{equation}
\b_1=\D-\frac{d}{2},\qquad\b_3=0
\end{equation}
hold. In the same way we find that the second equation takes the form
\begin{equation}
\label{otwo}
(p_2^2-p_4^2)\bigg((d-2\D+2\b_2)\,I_{\tilde\a+1\{\b_1,\b_2-1,\b_3\}}-2\b_3\,I_{\tilde\a+1\{\b_1,\b_2,\b_3-1\}}\bigg)=0
\end{equation}
and it is satisfied if 
\begin{equation}
\b_2=\D-\frac{d}{2},\qquad \b_3=0.
\end{equation}
One can check that the third equation 
\begin{equation}
p_2^2(d-2\D+2\b_2)\,I_{\tilde\a+1\{\b_1,\b_2-1,\b_3\}}-p_1^2(d-2\D+2\b_1)\,I_{\tilde\a+1\{\b_1-1,\b_2,\b_3\}}-2(p_2^2-p_1^2)\b_3\,I_{\tilde\a+1\{\b_1,\b_2,\b_3-1\}}=0,
\end{equation}
generates the same conditions given by \eqref{oone} and \eqref{otwo}.
After some computations, finally the solution for the  4-point function, in this particular case, can be written as
\begin{equation}
\braket{O(p_1)\,O(p_2)\,O(p_3)\,O(\bar{p}_4)}=\,\bar{\a} \,I_{\frac{d}{2}-1\left\{\D-\frac{d}{2},\D-\frac{d}{2},0\right\}}(p_1\, p_3;p_2\,p_4; s\,t),
\end{equation}
where $\bar{\a}$ is an undetermined constant. 

In the case $\D_1=\D_3=\D_x$ and $\D_2=\D_4=\D_y$, the special CWI's can be written as
\begin{equation}
\left\{\begin{aligned}&\\[-1.9ex]
&\bigg [\frac{\partial^2}{\partial p_1^2}+\frac{(d-2\D_x+1)}{p_1}\frac{\partial}{\partial p_1}-\frac{\partial^2}{\partial p_3^2}-\frac{(d-2\D_x+1)}{p_3}\frac{\partial}{\partial p_3}+\frac{(p_1^2-p_3^2)}{st}\frac{\partial^2}{\partial s\partial t}\bigg ]\,I_{\tilde\a\{\b_1,\b_2,\b_3\}}=0\\[1ex]
&\bigg[\frac{\partial^2}{\partial p_2^2}+\frac{(d-2\D_y+1)}{p_2}\frac{\partial}{\partial p_2}-\frac{\partial^2}{\partial p_4^2}-\frac{(d-2\D_y+1)}{p_4}\frac{\partial}{\partial p_4}+\frac{(p_2^2-p_4^2)}{st}\frac{\partial^2}{\partial s\partial t}\bigg ]\,I_{\tilde\a\{\b_1,\b_2,\b_3\}}=0\\[1ex]
&\bigg[\frac{\partial^2}{\partial p_3^2}+\frac{(d-2\D_x+1)}{p_3}\frac{\partial}{\partial p_3}-\frac{\partial^2}{\partial p_4^2}-\frac{(d-2\D_y+1)}{p_4}\frac{\partial}{\partial p_4}+\frac{(p_2^2-p_1^2)}{st}\frac{\partial^2}{\partial s\partial t}\bigg ]\,I_{\tilde\a\{\b_1,\b_2,\b_3\}}=0\\[1.3ex]
\end{aligned}\right.
\end{equation}
whose solution is
\begin{align}
\label{ssm1}
\braket{O(p_1)\,O(p_2)\,O(p_3)\,O(\bar{p}_4)}&=\,\bar{\bar{\a}} \,I_{\frac{d}{2}-1\left\{\D_x-\frac{d}{2},\D_y-\frac{d}{2},0\right\}}(p_1\, p_3;p_2\,p_4; s\,t)
\end{align}
which takes a form similar to the one typical of the three-point function given in \eqref{caz}.\\
In order to identify the form of the unique solution we need to satisfy the symmetry constraints and the absence of unphysical singularities \cite{2014JHEP...03..111B} in the domain of convergence. We will address the first issue below, while the second is discussed in \secref{convergence}, where we show that such singularities are not present. 

\subsection{Symmetric solutions as \texorpdfstring{$F_4$}{F4} hypergeometrics or 3K integrals. The equal scalings case}
The derivation of symmetric expressions of such correlators requires some effort, and can be obtained either by 
using the few known relations available for the Appell function $F_4$ or, alternatively (and more effectively), by resorting to the formalism of the 3K integrals. \\ 
A solution which is symmetric respect to all the permutation of the momenta $p_i$, expressed in terms of 3 of the four constants $c(a,b)$, after some manipulations, can be expressed in the form
\begin{align}
\braket{O(p_1)O(p_2)O(p_3)O(p_4)}&=\notag\\
&\hspace{-2cm}=\sum_{a,b}c(a,b)\Bigg[\,(s^2\,t^2)^{\D-\frac{3}{4}d}\left(\frac{p_1^2p_3^2}{s^2t^2}\right)^a\left(\frac{p_2^2p_4^2}{s^2t^2}\right)^bF_4\left(\a(a,b),\b(a,b),\g(a),\g'(b),\frac{p_1^2p_3^2}{s^2t^2},\frac{p_2^2p_4^2}{s^2t^2}\right)\notag\\
&\hspace{-1.5cm}+\,(s^2\,u^2)^{\D-\frac{3}{4}d}\,\left(\frac{p_2^2p_3^2}{s^2u^2}\right)^{a}\left(\frac{p_1^2p_4^2}{s^2u^2}\right)^{b}F_4\left(\a(a,b),\b(a,b),\g(a),\g'(b),\frac{p_2^2p_3^2}{s^2u^2},\frac{p_1^2p_4^2}{s^2u^2}\right)\notag\\
&\hspace{-1.5cm}+\,(t^2\,u^2)^{\D-\frac{3}{4}d}\,\left(\frac{p_1^2p_2^2}{t^2u^2}\right)^{a}\,\left(\frac{p_3^2p_4^2}{t^2u^2}\right)^{b}\,F_4\left(\a(a,b),\b(a,b),\g(a),\g'(b),\frac{p_1^2p_2^2}{t^2u^2},\frac{p_3^2p_4^2}{t^2u^2}\right)
\Bigg]
\label{fform}
\end{align}
where the four coefficients $c(a,b)$'s given in \eqref{solution} are reduced to three by the constraint 
\begin{align}
c\left(0,\D-\frac{d}{2}\right)=c\left(\D-\frac{d}{2},0\right).
\end{align}
Additional manipulations, in order to reduce even further the integration constants are hampered by absence of known 
relations for the Appell functions $F_4$. As already mentioned above, it is possible, though, to bypass the problem by turning to the 3K formalism. Equation \eqref{fform} can be further simplified using this formalism. 

\subsubsection{3K symmetrization in the equal scaling case}
In order to show this, \eqref{fform} can be written in terms of a linear combination of 3K integrals as
\begin{align}
\braket{O(p_1)O(p_2)O(p_3)O(p_4)}&= C_1\,I_{\frac{d}{2}-1\{\D-\frac{d}{2},\D-\frac{d}{2},0\}}(p_1\,p_3,p_2\,p_4,s\,t)\notag\\[1.2ex]
&\hspace{-1.5cm}+C_2\, \,I_{\frac{d}{2}-1\{\D-\frac{d}{2},\D-\frac{d}{2},0\}}(p_2\,p_3,p_1\,p_4,s\,u)+C_3\, \,I_{\frac{d}{2}-1\{\D-\frac{d}{2},\D-\frac{d}{2},0\}}(p_1\,p_2,p_3\,p_4,t\,u)
\end{align}
by an explicit symmetrization of the momenta in the parametric integrals.  It is now much simpler to show that the symmetry under permutations forces the $C_i$ to take the same value, and the final symmetric result is given by
\begin{align}
\braket{O(p_1)O(p_2)O(p_3)O(p_4)}&= C\bigg[\,I_{\frac{d}{2}-1\{\D-\frac{d}{2},\D-\frac{d}{2},0\}}(p_1\,p_3,p_2\,p_4,s\,t)\notag\\
&\hspace{-1.5cm}+\, \,I_{\frac{d}{2}-1\{\D-\frac{d}{2},\D-\frac{d}{2},0\}}(p_2\,p_3,p_1\,p_4,s\,u)+\, \,I_{\frac{d}{2}-1\{\D-\frac{d}{2},\D-\frac{d}{2},0\}}(p_1\,p_2,p_3\,p_4,t\,u)\bigg],
\end{align}
written in terms of only one arbitrary overall constant $C$. We can use the relation between the 3K integrals and the $F_4$ written in \eqref{3K} and \eqref{3Kplus},  to re-express the final symmetric solution, originally given in Eq.\eqref{fform}, in terms of a single constant in the form
\begin{align}
\braket{O(p_1)O(p_2)O(p_3)O(p_4)}&=2^{\frac{d}{2}-4}\ \ C\,\sum_{\l,\m=0,\D-\frac{d}{2}}\x(\l,\m)\bigg[\big(s^2\,t^2\big)^{\D-\frac{3}{4}d}\left(\frac{p_1^2 p_3^2}{s^2 t^2}\right)^\l\left(\frac{p_2^2p_4^2}{s^2t^2}\right)^\m\nonumber\\
&\hspace{-3cm}\times\,F_4\left(\frac{3}{4}d-\D+\l+\m,\frac{3}{4}d-\D+\l+\m,1-\D+\frac{d}{2}+\l,1-\D+\frac{d}{2}+\m,\frac{p_1^2 p_3^2}{s^2 t^2},\frac{p_2^2 p_4^2}{s^2 t^2}\right)\notag\\
&+\big(s^2\,u^2\big)^{\D-\frac{3}{4}d}\left(\frac{p_2^2 p_3^2}{s^2 u^2}\right)^\l\left(\frac{p_1^2p_4^2}{s^2u^2}\right)^\m\notag\\
&\hspace{-3cm}\times\,F_4\left(\frac{3}{4}d-\D+\l+\m,\frac{3}{4}d-\D+\l+\m,1-\D+\frac{d}{2}+\l,1-\D+\frac{d}{2}+\m,\frac{p_2^2 p_3^2}{s^2 u^2},\frac{p_1^2 p_4^2}{s^2 u^2}\right)\notag\\
&+\big(t^2\,u^2\big)^{\D-\frac{3}{4}d}\left(\frac{p_1^2 p_2^2}{t^2 u^2}\right)^\l\left(\frac{p_3^2p_4^2}{t^2u^2}\right)^\m\notag\\
&\hspace{-3cm}\times\,F_4\left(\frac{3}{4}d-\D+\l+\m,\frac{3}{4}d-\D+\l+\m,1-\D+\frac{d}{2}+\l,1-\D+\frac{d}{2}+\m,\frac{p_1^2 p_2^2}{t^2 u^2},\frac{p_3^2 p_4^2}{t^2 u^2}\right)\bigg].\label{finalSol}
\end{align}
where the coefficients $\x(\l,\m)$ are explicitly given by
\begin{equation}
\begin{split}
\x\left(0,0\right)&=\left[\G\left(\frac{3}{4}d-\D\right)\right]^2\left[\G\left(\D-\frac{d}{2}\right)\right]^2\\
\x\left(0,\D-\frac{d}{2}\right)&=\x\left(\D-\frac{d}{2},0\right)=\left[\G\left(\frac{d}{4}\right)\right]^2\G\left(\D-\frac{d}{2}\right)\G\left(\frac{d}{2}-\D\right)\\
\x\left(\D-\frac{d}{2},\D-\frac{d}{2}\right)&=\left[\G\left(\D-\frac{d}{4}\right)\right]^2\left[\G\left(\frac{d}{2}-\D\right)\right]^2.
\end{split}\label{xicoef}
\end{equation}
The solution found in \eqref{finalSol} is explicitly symmetric under all the possible permutations of the momenta and it is fixed up to one undetermined constant $C$. Eq. \eqref{finalSol} gives the final expression of the solution obtained from the first DCA \eqref{ansatz2}.

\section{Solutions from other DCA's }
The DCA from which we start is clearly not unique, since other types of factorized ans\"atze can be chosen in dual coordinate space. It is then resonable to ask whether the types of solutions that we have identified are truly unique, even if they are generated starting from a specific DCA. In order to answer such a question we turn to a different DCA and show that this is indeed the case. The intermediate steps of the derivation are rather involved, but one can obtain the same expression of the DCC solution obtained from \eqref{ansatz2}, given in \eqref{finalSol}, using some analytic continuations of the new solution generated by such a second ans\"atz. \\
For this purpose, we consider as a starting point a DCA of the form 
\begin{equation}
\Phi=\left(p_1^2\,p_3^2\right)^{n_s}\,F\left(\frac{s^2t^2}{p_1^2p_3^2},\frac{p_2^2p_4^2}{p_1^2p_3^2}\right)
\label{ansatz3}
\end{equation}
where all the scalings are taken to be equal $\D_i=\D$, $i=1,2,3,4$. Also in this case
the dilatation WI's fix the value of $n_s$ as in \eqref{scaled}, while the special WI's can be written as
\begin{equation}
\resizebox{1\hsize}{!}{$
\left\{
\begin{aligned}\\[-1.6ex]
&\left[x(1-x)\partial_{xx}-2x y\partial_{xy}-y^2\partial_{yy}-y(d-\D+1)\partial_y+[1-(d-\D+1)x]\partial_x-\frac{d}{4}\left(\frac{3}{4}d-\D\right)\right] F(x,y)=0\\[1.5ex]
&\left[x\partial_{xx}-y\partial_{yy}+\partial_x-\left(\frac{d}{2}-\D+1\right)\partial_y\right]F(x,y)=0\\[1.3ex]
\end{aligned}\right.$ }
\end{equation}
where we have defined $x=s^2t^2/(p_1^2p_3^2)$ and $y=p_2^2p_4^2/(p_1^2p_3^2)$. Subtracting the second equation from the first one we derive the system of equations
\begin{equation}
\resizebox{1\hsize}{!}{$
\left\{
\begin{aligned}\\[-1.6ex]
&\left[x(1-x)\partial_{xx}-2x y\,\partial_{xy}-y^2\partial_{yy}-y(d-\D+1)\partial_y+[1-(d-\D+1)x]\partial_x-\frac{d}{4}\left(\frac{3}{4}d-\D\right)\right] F(x,y)=0\\[1.5ex]
&\Bigg[y(1-y)\partial_{yy}-2x y\,\partial_{xy}-x^2\partial_{xx}\\[-1ex]
&\hspace{2cm}-x(d-\D+1)\partial_x+\left[\left(\frac{d}{2}-\D+1\right)-(d-\D+1)x\right]\partial_x-\frac{d}{4}\left(\frac{3}{4}d-\D\right)\Bigg] F(x,y)=0\\[1.3ex]
\end{aligned}\right. $}
\end{equation}
which corresponds, once more, to a hypergeometric system of equations in two variables, corresponding to Appell's $F_4$. The general solution of such a system can be expressed as a linear combination of two $F_4$ functions as
\begin{align}
\Phi&=\left(p_1^2\,p_3^2\right)^{\D-\frac{3}{4}d}\bigg[C_1 F_4\left(\frac{d}{4}\,,\,\frac{3}{4}d-\D\, ,\,1\,,\,\frac{d}{2}-\D+1\,;\frac{s^2t^2}{p_1^2p_3^2}\,,\,\frac{p_2^2p_4^2}{p_1^2p_3^2}\right)\notag\\
&\hspace{2.5cm}+C_2\left(\frac{p_2^2p_4^2}{p_1^2p_3^2}\right)^{\D-\frac{d}{2}} F_4\left(\D-\frac{d}{4}\,,\,\frac{d}{4}\, ,\,1\,,\,1-\frac{d}{2}+\D\,;\frac{s^2t^2}{p_1^2p_3^2}\,,\,\frac{p_2^2p_4^2}{p_1^2p_3^2}\right)\bigg].\label{Solution1}
\end{align}
This solution corresponds to a very specific case, in which one of the 4 parameters of the general solution given by the 4 hypergeometric functions of type $F_4$ is fixed to $\g=1$. One can show that in this case the number of independent hypergeometric solutions is then reduced from 4 to 2.  However, at this stage, Eq. \eqref{Solution1} is symmetric only respect to the momentum exchanges $(p_1\leftrightarrow p_3)$ and $(p_2\leftrightarrow p_4)$. 
As a first step we can proceed by constructing the completely symmetric solution of the same system in the form
\begin{align}
\Phi&=\left(p_1^2\,p_3^2\right)^{\D-\frac{3}{4}d}\bigg[C_1 F_4\left(\frac{d}{4}\,,\,\frac{3}{4}d-\D\, ,\,1\,,\,\frac{d}{2}-\D+1\,;\frac{s^2t^2}{p_1^2p_3^2}\,,\,\frac{p_2^2p_4^2}{p_1^2p_3^2}\right)\notag\\
&\hspace{2.5cm}+C_2\left(\frac{p_2^2p_4^2}{p_1^2p_3^2}\right)^{\D-\frac{d}{2}} F_4\left(\D-\frac{d}{4}\,,\,\frac{d}{4}\, ,\,1\,,\,1-\frac{d}{2}+\D\,;\frac{s^2t^2}{p_1^2p_3^2}\,,\,\frac{p_2^2p_4^2}{p_1^2p_3^2}\right)\bigg]\notag\\
&+\left(p_2^2\,p_3^2\right)^{\D-\frac{3}{4}d}\bigg[C_1 F_4\left(\frac{d}{4}\,,\,\frac{3}{4}d-\D\, ,\,1\,,\,\frac{d}{2}-\D+1\,;\frac{s^2u^2}{p_2^2p_3^2}\,,\,\frac{p_1^2p_4^2}{p_2^2p_3^2}\right)\notag\\
&\hspace{2.5cm}+C_2\left(\frac{p_1^2p_4^2}{p_2^2p_3^2}\right)^{\D-\frac{d}{2}} F_4\left(\D-\frac{d}{4}\,,\,\frac{d}{4}\, ,\,1\,,\,1-\frac{d}{2}+\D\,;\frac{s^2u^2}{p_2^2p_3^2}\,,\,\frac{p_1^2p_4^2}{p_2^2p_3^2}\right)\bigg]\notag\\
&+\left(p_1^2\,p_2^2\right)^{\D-\frac{3}{4}d}\bigg[C_1 F_4\left(\frac{d}{4}\,,\,\frac{3}{4}d-\D\, ,\,1\,,\,\frac{d}{2}-\D+1\,;\frac{u^2t^2}{p_1^2p_2^2}\,,\,\frac{p_3^2p_4^2}{p_1^2p_2^2}\right)\notag\\
&\hspace{2.5cm}+C_2\left(\frac{p_3^2p_4^2}{p_1^2p_2^2}\right)^{\D-\frac{d}{2}} F_4\left(\D-\frac{d}{4}\,,\,\frac{d}{4}\, ,\,1\,,\,1-\frac{d}{2}+\D\,;\frac{u^2t^2}{p_1^2p_2^2}\,,\,\frac{p_3^2p_4^2}{p_1^2p_2^2}\right)\bigg]\label{Phi}
\end{align}
containing only the coefficients $C_1$ and $C_2$.  Considering the $(p_2\leftrightarrow p_4)$ exchange the solution will be given by
\begin{align}
\Phi_{p_2\leftrightarrow p_4}&=\left(p_1^2\,p_3^2\right)^{\D-\frac{3}{4}d}\bigg[C_1 F_4\left(\frac{d}{4}\,,\,\frac{3}{4}d-\D\, ,\,1\,,\,\frac{d}{2}-\D+1\,;\frac{s^2t^2}{p_1^2p_3^2}\,,\,\frac{p_2^2p_4^2}{p_1^2p_3^2}\right)\notag\\
&\hspace{2.5cm}+C_2\left(\frac{p_2^2p_4^2}{p_1^2p_3^2}\right)^{\D-\frac{d}{2}} F_4\left(\D-\frac{d}{4}\,,\,\frac{d}{4}\, ,\,1\,,\,1-\frac{d}{2}+\D\,;\frac{s^2t^2}{p_1^2p_3^2}\,,\,\frac{p_2^2p_4^2}{p_1^2p_3^2}\right)\bigg]\notag\\
&+\left(p_4^2\,p_3^2\right)^{\D-\frac{3}{4}d}\bigg[C_1 F_4\left(\frac{d}{4}\,,\,\frac{3}{4}d-\D\, ,\,1\,,\,\frac{d}{2}-\D+1\,;\frac{t^2u^2}{p_4^2p_3^2}\,,\,\frac{p_1^2p_2^2}{p_4^2p_3^2}\right)\notag
\end{align}
\begin{align}
&\hspace{2.5cm}+C_2\left(\frac{p_1^2p_4^2}{p_2^2p_3^2}\right)^{\D-\frac{d}{2}} F_4\left(\D-\frac{d}{4}\,,\,\frac{d}{4}\, ,\,1\,,\,1-\frac{d}{2}+\D\,;\frac{t^2u^2}{p_4^2p_3^2}\,,\,\frac{p_1^2p_2^2}{p_4^2p_3^2}\right)\bigg]\notag\\
&+\left(p_1^2\,p_4^2\right)^{\D-\frac{3}{4}d}\bigg[C_1 F_4\left(\frac{d}{4}\,,\,\frac{3}{4}d-\D\, ,\,1\,,\,\frac{d}{2}-\D+1\,;\frac{u^2s^2}{p_1^2p_4^2}\,,\,\frac{p_3^2p_2^2}{p_1^2p_4^2}\right)\notag\\
&\hspace{2.5cm}+C_2\left(\frac{p_3^2p_2^2}{p_1^2p_4^2}\right)^{\D-\frac{d}{2}} F_4\left(\D-\frac{d}{4}\,,\,\frac{d}{4}\, ,\,1\,,\,1-\frac{d}{2}+\D\,;\frac{u^2s^2}{p_1^2p_4^2}\,,\,\frac{p_3^2p_2^2}{p_1^2p_4^2}\right)\bigg]
\end{align}
that can be rearranged in the form given in \eqref{Phi} using \eqref{transfF4}.\\
After imposing the symmetry condition $\Phi_{p_2\leftrightarrow p_4}=\Phi$ under this particular permutation, we find a single degenerate condition on the ratios of $C_1$ and $C_2$ given by
\begin{equation}
\resizebox{1\hsize}{!}{$\frac{C_1}{C_2}=\left[\G\left(\D-\frac{3}{4}d\right)\G\left(1-\D+\frac{3}{4}d\right)\G\left(1+\D-\frac{d}{2}\right)\right]\left[\G\left(\D-\frac{d}{4}\right)\G\left(1+\D-\frac{3}{4}d\right)\G\left(1-\D+\frac{d}{2}\right)\right]^{-1}.$}
\end{equation}
This constraint fixes the solution up to one undetermined constant in the form
\begin{equation}
\resizebox{1\hsize}{!}{$
\begin{aligned}
\Phi&=C_1\bigg\{\left(p_1^2\,p_3^2\right)^{\D-\frac{3}{4}d}\bigg[F_4\left(\frac{d}{4}\,,\,\frac{3}{4}d-\D\,,\,1\,,\,\frac{d}{2}-\D+1\,;\frac{s^2t^2}{p_1^2p_3^2}\,,\,\frac{p_2^2p_4^2}{p_1^2p_3^2}\right)\\[1.2ex]
&\hspace{-0.3cm}+\frac{\G\left(\D-\frac{d}{4}\right)\G\left(1+\D-\frac{3}{4}d\right)\G\left(1-\D+\frac{d}{2}\right)}{\G\left(\D-\frac{3}{4}d\right)\G\left(1-\D+\frac{3}{4}d\right)\G\left(1+\D-\frac{d}{2}\right)}\left(\frac{p_2^2p_4^2}{p_1^2p_3^2}\right)^{\D-\frac{d}{2}} F_4\left(\D-\frac{d}{4}\,,\,\frac{d}{4}\, ,\,1\,,\,1-\frac{d}{2}+\D\,;\frac{s^2t^2}{p_1^2p_3^2}\,,\,\frac{p_2^2p_4^2}{p_1^2p_3^2}\right)\bigg]\\[1.2ex]
&+\left(p_2^2\,p_3^2\right)^{\D-\frac{3}{4}d}\bigg[F_4\left(\frac{d}{4}\,,\,\frac{3}{4}d-\D\, ,\,1\,,\,\frac{d}{2}-\D+1\,;\frac{s^2u^2}{p_2^2p_3^2}\,,\,\frac{p_1^2p_4^2}{p_2^2p_3^2}\right)\\[1.2ex]
&\hspace{-0.3cm}+\frac{\G\left(\D-\frac{d}{4}\right)\G\left(1+\D-\frac{3}{4}d\right)\G\left(1-\D+\frac{d}{2}\right)}{\G\left(\D-\frac{3}{4}d\right)\G\left(1-\D+\frac{3}{4}d\right)\G\left(1+\D-\frac{d}{2}\right)} \left(\frac{p_1^2p_4^2}{p_2^2p_3^2}\right)^{\D-\frac{d}{2}} F_4\left(\D-\frac{d}{4}\,,\,\frac{d}{4}\, ,\,1\,,\,1-\frac{d}{2}+\D\,;\frac{s^2u^2}{p_2^2p_3^2}\,,\,\frac{p_1^2p_4^2}{p_2^2p_3^2}\right)\bigg]\\[1.2ex]
&+\left(p_1^2\,p_2^2\right)^{\D-\frac{3}{4}d}\bigg[F_4\left(\frac{d}{4}\,,\,\frac{3}{4}d-\D\, ,\,1\,,\,\frac{d}{2}-\D+1\,;\frac{u^2t^2}{p_1^2p_2^2}\,,\,\frac{p_3^2p_4^2}{p_1^2p_2^2}\right)\\[1.2ex]
&\hspace{-0.3cm}+\frac{\G\left(\D-\frac{d}{4}\right)\G\left(1+\D-\frac{3}{4}d\right)\G\left(1-\D+\frac{d}{2}\right)}{\G\left(\D-\frac{3}{4}d\right)\G\left(1-\D+\frac{3}{4}d\right)\G\left(1+\D-\frac{d}{2}\right)}\left(\frac{p_3^2p_4^2}{p_1^2p_2^2}\right)^{\D-\frac{d}{2}} F_4\left(\D-\frac{d}{4}\,,\,\frac{d}{4}\, ,\,1\,,\,1-\frac{d}{2}+\D\,;\frac{u^2t^2}{p_1^2p_2^2}\,,\,\frac{p_3^2p_4^2},{p_1^2p_2^2}\right)\bigg]\bigg\}\label{finalsolution}
\end{aligned}$}
\end{equation}
which can be shown to be symmetric under all the possible permutations of the momenta $(p_1,\,p_2,\,p_3,\,p_4)$.\\
We are now going to show the equivalence of such solution to \eqref{finalSol}, which is given by a 3K integral. We perform an analytic continuation of \eqref{finalsolution} using \eqref{transfF4} to obtain the intermediate expression
\begin{equation}
\resizebox{1\hsize}{!}{$
\begin{aligned}
\Phi&=\left(s^2\,t^2\right)^{\D-\frac{3}{4}d}C_1\Bigg[ \frac{\G\left(\D-\frac{d}{2}\right)(-1)^{\D-\frac{3}{4}d}}{\G\left(1+\D-\frac{3}{4}d\right)\G\left(\frac{d}{4}\right)}\,F_4\left(\frac{3}{4}d-\D\,,\,\frac{3}{4}d-\D\, ,\,\frac{d}{2}-\D+1\,,\,\frac{d}{2}-\D+1\,;\,\frac{p_1^2p_3^2}{s^2t^2}\, ,\,\frac{p_2^2p_4^2}{s^2t^2}\right)\\[1.2ex]
&\hspace{2.5cm}+\frac{\G\left(\frac{d}{2}-\D\right)(-1)^{-\frac{d}{4}}}{\G\left(1-\frac{d}{4}\right)\G\left(\frac{3}{4}d-
	\D\right)}\left(\frac{p_1^2p_3^2}{s^2t^2}\right)^{\D-\frac{d}{2}} F_4\left(\frac{d}{4}\,,\,\frac{d}{4}\, ,\,1-\frac{d}{2}+\D\,,\,\,1+\frac{d}{2}-\D\,;\,\frac{p_1^2p_3^2}{s^2t^2},\frac{p_2^2p_4^2}{s^2t^2}\right)\\[1.2ex]
&\hspace{2.5cm}+\frac{\G\left(\frac{d}{2}-\D\right)(-1)^{-\frac{d}{4}}}{\G\left(1-\frac{d}{4}\right)\G\left(\frac{3}{4}d-
	\D\right)}\left(\frac{p_2^2p_4^2}{s^2t^2}\right)^{\D-\frac{d}{2}} F_4\left(\frac{d}{4}\,,\,\frac{d}{4}\, ,\,1+\frac{d}{2}-\D\,,\,\,1-\frac{d}{2}+\D\,;\,\frac{p_1^2p_3^2}{s^2t^2},\frac{p_2^2p_4^2}{s^2t^2}\right)\\[1.2ex]
&\hspace{2.5cm}+\frac{\left[\G\left(\frac{d}{2}-\D\right)\right]^2\G\left(\D-\frac{d}{4}\right)(-1)^{\frac{d}{4}-\D}}{\G\left(1+\frac{d}{4}-\D\right)\G\left(\frac{d}{4}\right)\G\left(\frac{3}{4}d-\D\right)\G\left(\D-\frac{d}{2}\right)} \\
&\hspace{2cm}\times \left(\frac{p_2^2p_4^2}{s^2t^2}\right)^{\D-\frac{d}{2}}\left(\frac{p_1^2p_3^2}{s^2t^2}\right)^{\D-\frac{d}{2}}F_4\left(\frac{3}{4}d-\D\,,\,\frac{3}{4}d-\D\, ,\,1+\frac{d}{2}-\D\,,\,\,1+\frac{d}{2}-\D\,;\,\frac{p_1^2p_3^2}{s^2t^2},\frac{p_2^2p_4^2}{s^2t^2}\right)\Bigg] \\[1.2ex]
&\hspace{2cm} +[(p_1\leftrightarrow p_2)]+[(p_2\leftrightarrow p_3)].\label{final}
\end{aligned}$}
\end{equation}
After some manipulations and using the properties of Gamma function
\begin{equation}
\G(a-b)=\frac{\G(a)\G(1-a)(-1)^b}{\G(1-a+b)},\hspace{1cm}\frac{1}{\G(a-b)}=\frac{\G(1-a+b)(-1)^{-b}}{\G(a)\G(1-a)},
\end{equation}
we write \eqref{final} as
\begin{align}
\Phi&=\left(s^2\,t^2\right)^{\D-\frac{3}{4}d}\frac{C_1}{\G\left(\frac{d}{4}\right)\G\left(\frac{d}{2}\right)\G\left(1-\frac{d}{2}\right)\G\left(\frac{3}{4}d-\D\right)\G\left(\D-\frac{d}{2}\right)}\notag\\[1.2ex]
&\hspace{-0.5cm}\times\Bigg\{ \left[\G\left(\D-\frac{d}{2}\right)\right]^2\left[\G\left(\frac{3}{4}d-\D\right)\right]^2\,F_4\left(\frac{3}{4}d-\D\,,\,\frac{3}{4}d-\D\, ,\,\frac{d}{2}-\D+1\,,\,\frac{d}{2}-\D+1\,;\,\frac{p_1^2p_3^2}{s^2t^2}\, ,\,\frac{p_2^2p_4^2}{s^2t^2}\right)\notag
\end{align}
\begin{align}
&\hspace{-0.5cm}+\left[\G\left(\frac{d}{4}\right)\right]^2\G\left(\frac{d}{2}-\D\right)\G\left(\D-\frac{d}{2}\right)\left(\frac{p_1^2p_3^2}{s^2t^2}\right)^{\D-\frac{d}{2}}  F_4\left(\frac{d}{4}\,,\,\frac{d}{4}\, ,\,1-\frac{d}{2}+\D\,,\,\,1+\frac{d}{2}-\D\,;\,\frac{p_1^2p_3^2}{s^2t^2},\frac{p_2^2p_4^2}{s^2t^2}\right)\notag\\[1.2ex]
&\hspace{-0.5cm}+\left[\G\left(\frac{d}{4}\right)\right]^2\G\left(\frac{d}{2}-\D\right)\G\left(\D-\frac{d}{2}\right)\left(\frac{p_2^2p_4^2}{s^2t^2}\right)^{\D-\frac{d}{2}} F_4\left(\frac{d}{4}\,,\,\frac{d}{4}\, ,\,1+\frac{d}{2}-\D\,,\,\,1-\frac{d}{2}+\D\,;\,\frac{p_1^2p_3^2}{s^2t^2},\frac{p_2^2p_4^2}{s^2t^2}\right)\notag\\[1.2ex]
&\hspace{-0.5cm}+\left[\G\left(\frac{d}{2}-\D\right)\right]^2\left[\G\left(\D-\frac{d}{4}\right)\right]^2 \notag\\
&\hspace{1cm}\times \left(\frac{p_2^2p_4^2}{s^2t^2}\right)^{\D-\frac{d}{2}}\left(\frac{p_1^2p_3^2}{s^2t^2}\right)^{\D-\frac{d}{2}}F_4\left(\frac{3}{4}d-\D\,,\,\frac{3}{4}d-\D\, ,\,1+\frac{d}{2}-\D\,,\,\,1+\frac{d}{2}-\D\,;\,\frac{p_1^2p_3^2}{s^2t^2},\frac{p_2^2p_4^2}{s^2t^2}\right)\Bigg\}\notag\\[1.2ex]
&\hspace{2cm} +[(p_1\leftrightarrow p_2)]+[(p_2\leftrightarrow p_3)].
\end{align}
which takes the same form of solution \eqref{finalSol}. In fact this expression can be rewritten as
\begin{align}
\Phi&=\frac{C_1}{\G\left(\frac{d}{4}\right)\G\left(\frac{d}{2}\right)\G\left(1-\frac{d}{2}\right)\G\left(\frac{3}{4}d-\D\right)\G\left(\D-\frac{d}{2}\right)}\,\sum_{\l,\m=0,\D-\frac{d}{2}}\x(\l,\m)\bigg[\big(s^2\,t^2\big)^{\D-\frac{3}{4}d}\left(\frac{p_1^2 p_3^2}{s^2 t^2}\right)^\l\left(\frac{p_2^2p_4^2}{s^2t^2}\right)^\m\notag\\
&\hspace{1cm}\times\,F_4\left(\frac{3}{4}d-\D+\l+\m,\frac{3}{4}d-\D+\l+\m,1-\D+\frac{d}{2}+\l,1-\D+\frac{d}{2}+\m,\frac{p_1^2 p_3^2}{s^2 t^2},\frac{p_2^2 p_4^2}{s^2 t^2}\right)\notag\\
&+\big(s^2\,u^2\big)^{\D-\frac{3}{4}d}\left(\frac{p_2^2 p_3^2}{s^2 u^2}\right)^\l\left(\frac{p_1^2p_4^2}{s^2u^2}\right)^\m\notag\\
&\hspace{1cm}\times\,F_4\left(\frac{3}{4}d-\D+\l+\m,\frac{3}{4}d-\D+\l+\m,1-\D+\frac{d}{2}+\l,1-\D+\frac{d}{2}+\m,\frac{p_2^2 p_3^2}{s^2 u^2},\frac{p_1^2 p_4^2}{s^2 u^2}\right)\notag\\
&+\big(t^2\,u^2\big)^{\D-\frac{3}{4}d}\left(\frac{p_1^2 p_2^2}{t^2 u^2}\right)^\l\left(\frac{p_3^2p_4^2}{t^2u^2}\right)^\m\notag\\
&\hspace{1cm}\times\,F_4\left(\frac{3}{4}d-\D+\l+\m,\frac{3}{4}d-\D+\l+\m,1-\D+\frac{d}{2}+\l,1-\D+\frac{d}{2}+\m,\frac{p_1^2 p_2^2}{t^2 u^2},\frac{p_3^2 p_4^2}{t^2 u^2}\right)\bigg]
\label{finf}
\end{align}
with a different coefficient in front, but with the coefficients $\xi(\l,\m)$ being the same of \eqref{xicoef}, completing the proof. \\
Notice that we could have gone through the analytic proof of the equivalence, by using even a third DCA, for instance of the form 
\begin{equation}
\Phi'=\left(p_2^2\,p_4^2\right)^{n'_s}\,F\left(\frac{s^2t^2}{p_2^2p_4^2},\frac{p_1^2p_3^2}{p_2^2p_4^2}\right)
\end{equation}
and following the same procedure described above, we would have obtained an hypergeometric system of equations with a solution of the form
\begin{align}
\label{sol2}
\Phi'&=\left(p_2^2\,p_4^2\right)^{\D-\frac{3}{4}d}\bigg[C_1 F_4\left(\frac{d}{4}\,,\,\frac{3}{4}d-\D\, ,\,1\,,\,\frac{d}{2}-\D+1\,;\,\frac{p_1^2p_3^2}{p_2^2p_4^2}\, ,\,\frac{s^2t^2}{p_2^2p_4^2}\right)\notag\\
&\hspace{2.5cm}+C_2\left(\frac{p_1^2p_3^2}{p_2^2p_4^2}\right)^{\D-\frac{d}{2}} F_4\left(\D-\frac{d}{4}\,,\,\frac{d}{4}\, ,\,1\,,\,\,1-\frac{d}{2}+\D\,;\,\frac{p_1^2p_3^2}{p_2^2p_4^2},\frac{s^2t^2}{p_2^2p_4^2}\right)\bigg],
\end{align}
as in \eqref{Solution1}. It can be explicitly shown that also in this case, by repeating the steps illustrated above, from \eqref{sol2} one arrives to \eqref{finalSol}. \\
We have indeed shown that DCC solutions take a unique form, independently of the structure of the original DCA. If we combine the results of section \ref{dccsection} with those above, it is clear that the solutions that we have found represent DCC correlators for any spacetime dimensions, of which the box diagram and its melonic variants 
are the only perturbtive realization, limited to $d=4$. 

\section{Convergence of the 3K solution integral and absence of physical singularities}\label{convergence}
The absence of unphysical singularities in the domain of convergence of the solution found, given in \eqref{ssm1}, can be addressed as follows. \\
Considering the DCC solution, we have derived its explicit expression as
\begin{align}
&I_{\frac{d}{2}-1\{\Delta-\frac{d}{2},\Delta-\frac{d}{2},0\}}(p_1p_3,p_2p_4,s,t)=\notag\\
&\qquad=\,(p_1p_3)^{\Delta-\frac{d}{2}}(p_2p_4)^{\Delta-\frac{d}{2}}\int_{0}^\infty\,dx\,x^{\frac{d}{2}-1}\,K_{\Delta-\frac{d}{2}}(p_1p_3\,x)\,K_{\Delta-\frac{d}{2}}(p_2p_4\,x)\,K_{0}(st\,x).\label{Sol}
\end{align}
Notice that a possible singularity which could invalidate the convergence of \eqref{Sol} can be generated by the Bessel function $K_0(x)$ at small $x$, as evident from the expansions 
\begin{align}
\label{questoK}
&K_{\nu}(x)\simeq\,\sqrt{\frac{\pi}{2}}\,\frac{e^{-x}}{\sqrt{x}}+\dots&&\text{at large $x$},\\
&K_{\nu}(x)\simeq\,x^\nu\,\frac{\Gamma(-\nu)}{2^{1+\nu}}+x^{-\nu}\,\frac{\Gamma(\nu)}{2^{1-\nu}}+\dots&&\text{at small $x$}.
\end{align}
The singularity in $K_0$ can be regulated using the replacement 
$K_0\to K_\epsilon$, with $\epsilon$ a small regulator parameter $(\epsilon >0)$. For this purpose we consider the regulated expression
\begin{align}
&I_{\frac{d}{2}-1\{\Delta-\frac{d}{2},\Delta-\frac{d}{2},\epsilon\}}(p_1p_3,p_2p_4,s,t)=\notag\\
&\quad=\,(p_1p_3)^{\Delta-\frac{d}{2}}(p_2p_4)^{\Delta-\frac{d}{2}}(st)^\epsilon\int_{0}^\infty\,dx\,x^{\frac{d}{2}-1}\,K_{\Delta-\frac{d}{2}}(p_1p_3\,x)\,K_{\Delta-\frac{d}{2}}(p_2p_4\,x)\,K_{\epsilon}(st\,x).\label{sol}
\end{align}
With this regularization, at large $x$ the integrand  of \eqref{Sol} can be written as
\begin{align}
\,x^{\frac{d}{2}-1}\,K_{\Delta-\frac{d}{2}}(p_1p_3\,x)\,K_{\Delta-\frac{d}{2}}(p_2p_4\,x)\,K_{\epsilon}(st\,x)\simeq\,(\sqrt{p_1p_2}\sqrt{p_3p_4}\sqrt{st})^{-1}\left(\frac{\pi}{2}\right)^{\frac{3}{2}}\,x^{\frac{d-5}{2}}\,e^{-(p_1p_3+p_2p_4+st)x}
\end{align}
which is well-behaved in the asymptotic region in $x$ if the condition
\begin{align}
p_1p_3+p_2p_4+st>0
\label{thecond}
\end{align}
is satisfied.\\
Similarly, the same integrand at small $x$ gives 
\begin{align}
&\,x^{\frac{d}{2}-1}\,K_{\Delta-\frac{d}{2}}(p_1p_3\,x)\,K_{\Delta-\frac{d}{2}}(p_2p_4\,x)\,K_{\epsilon}(st\,x)\notag\\
&\simeq\,x^{\frac{d}{2}-1}\left((p_1p_3\,x)^{\Delta-\frac{d}{2}}\frac{1}{2^{1+\Delta-\frac{d}{2}}}\G\left(\frac{d}{2}-\Delta\right)+(p_1p_3\,x)^{\frac{d}{2}-\Delta}\frac{1}{2^{1+\frac{d}{2}-\Delta}}\G\left(\Delta-\frac{d}{2}\right)\right)\notag\\
&\times\left((p_2p_4\,x)^{\Delta-\frac{d}{2}}\frac{1}{2^{1+\Delta-\frac{d}{2}}}\G\left(\frac{d}{2}-\Delta\right)+(p_2p_4\,x)^{\frac{d}{2}-\Delta}\frac{1}{2^{1+\frac{d}{2}-\Delta}}\G\left(\Delta-\frac{d}{2}\right)\right)\left((st\,x)^\epsilon\,\frac{\Gamma(-\epsilon)}{2^{1+\epsilon}}+(st\,x)^{-\epsilon}\,\frac{\Gamma(\epsilon)}{2^{1-\epsilon}}\right) \label{expSmallx}
\end{align} 
and expanding the last factor in the previous expression - for small values of  the regulator $\epsilon$ - this takes the form 
\begin{align}
(st\,x)^\epsilon\,\frac{\Gamma(-\epsilon)}{2^{1+\epsilon}}+(st\,x)^{-\epsilon}\,\frac{\Gamma(\epsilon)}{2^{1-\epsilon}}\simeq -\log(st\,x)-\gamma+\log(2)+O(\epsilon),
\end{align}
By combining all the contributions,  \eqref{expSmallx} can be rewritten as
\begin{align}
&\,x^{\frac{d}{2}-1}\,K_{\Delta-\frac{d}{2}}(p_1p_3\,x)\,K_{\Delta-\frac{d}{2}}(p_2p_4\,x)\,K_{\epsilon}(st\,x)\simeq\log(st\, x)\,x^{\frac{d}{2}-1\pm\left(\Delta-\frac{d}{2}\right)\pm\left(\Delta-\frac{d}{2}\right)} +O(\epsilon)
\end{align}
which converges if the condition
\begin{align}
\frac{d}{2}-1\pm\left(\Delta-\frac{d}{2}\right)\pm\left(\Delta-\frac{d}{2}\right)>0
\label{scling}
\end{align}
is satisfied which branches into four possible constraints. 
One can check that the bound \eqref{thecond} is satisfied in the physical region 
\begin{align} 
p_1+p_2+p_3+p_4>0
\end{align}
since 
\begin{equation}
p_1 p_3+p_2p_4+s t>p_1+p_2+p_3+p_4>0
\end{equation}
and the convergence of the 3K representation is guaranteed if \eqref{scling} is satisfied.
The condition \eqref{scling} can generate in the physical region some divergences which need an appropriate regularization, as pointed out in \cite{Bzowski:2015yxv, Bzowski:2015pba} in the case of 3-point functions. A similar analysis of the singularities in view of the previous constraints is underway and the regularization procedure will be presented in a separate work.

\section{CWI's at fixed angle  and the Lauricella hypergeometric functions}
From this section on we turn to an analysis of another class of solutions of the CWI's, approximate in their character, which also show the hypergeometric nature of the system of equations derived from the CWI's, if we investigate such equations in a special kinematical limit. 

The hypergeometric nature of the CWI's can be shown if we resort to some approximations. \\
The second class of solutions that we are going to discuss are obtained by assuming particular asymptotic values of the $s$ and $t$ invariants. In this case the solution is generated by inspecting the contribution coming from the operatorial term $D_{s t} $ defined below in Eq. \eqref{dst}, which vanishes if it acts on a function of the form $\log(t/s)$. Such solution, for dimensional reason, is unique, and can be included in a factorized ans\"atz in order to generate a solution of the full equations. As we are going to show, the choice of such ans\"atz takes to solutions in which the dependence on the external mass invariants $p_i^2$ and the $s,t$ invariants are completely factorized and describe asymptotic solutions of the equations for large $s$ and $t$ invariants. In this case the rapidity variable $y=\log(t/s)$ can be associated with the behaviour of the correlator at fixed angle (i.e. with $s/t$ fixed and $O(1)$). The remaining part of the solutions, in this case, are expressed as a system of generalized hypergeometric (Lauricella) functions. We will show that such solutions can be expressed in terms of 4-K integrals, that we will define. \\   
For this purpose is helpful to identify several contributions in the expressions of the $C_i's$ given above taking $C_{13}$ as an example. Beside the operator $\textup{K}_i$ given by \eqref{Koper}, we define in general the operators
\begin{align}
&J_{ij}=p_i\frac{\partial}{\partial p_i}+p_j\frac{\partial}{\partial p_j}, \qquad \tilde{J}_{ik}=p_i\frac{\partial}{\partial p_i}-p_k\frac{\partial}{\partial p_k} ,
\label{firstref}\\[1.2ex]
&J_{ij,kl}=p_i\frac{\partial}{\partial p_i}+p_j\frac{\partial}{\partial p_j}-p_k\frac{\partial}{\partial p_k}-p_l\frac{\partial}{\partial p_l}=J_{ij}-J_{kl}
\end{align}
\begin{equation}
h_{ij,kl}=\D_i+\D_j-\D_k-\D_l,\qquad
D^{t}=\frac{1}{t}\frac{\partial}{\partial t},\qquad  D^{s}=\frac{1}{s}\frac{\partial}{\partial s}
\end{equation}
and
{\begin{equation}
\label{dst}
D^{s t}_{ij}\equiv\frac{(p_i^2-p_j^2)}{s t}\frac{\partial^2}{\partial s\partial t}=(p_i^2-p_j^2) D^s D^t.
\end{equation}
These notations turns necessary when discussing the contributions of the various operators appearing in the equations in a compact way, but we will also turn to their their original (extended) expressions in order to avoid using indices, whenever possible.

For instance, $C_{13}$ will take the form 
\begin{equation}
C_{13}=\left( \textup{K}_{13} + D^s J_{12,34} +h_{34,12}D^s + D^t J_{14,23} +h_{23,14}D^t + D_{13}^{s t}\right)\Phi=0,
\end{equation}
while $C_1$ will be given by 
\begin{equation}
C_1=\left(\textup{K}_{14} +D^s J_{12,34}+ h_{34,12}D^s +D_{23}^{st}\right)\Phi=0.
\label{intform}
\end{equation}
Using the definitons above, each equation can be characterized in terms of the set of operators $(K, J, h D, D^{s t})$. 
We recognize in $\textup{K}_{ij}$ the typical operators appearing in 3-point functions, which emerge when every form factor is 
expressed in terms of the three external mass invariant, with the $J_{ij,kj}$ vanishing when the scaling dimensions of the same invariants are suitably balanced. For instance, given a function of two variables $f(z_1,z_2)$, 
we will have 
\begin{equation}
J_{ij,kl}\, f\left(\frac{p_i^2}{p_j^2},\frac{p_k^2}{p_l^2}\right)=0 \qquad J_{ij,kl}\, f\left(p_i^2 p_k^2,p_j^2 p_l^2\right)=0
\end{equation}
and similar equations obtained by suitably exchanging $i,j,k,l$. If all the external invariants are grouped into a single 
variable, for a given function $g(z)$, similarly we will obtain, for instance,
\begin{equation}
J_{ij} \,g\left(\frac{p_i^2}{ p_j^2}\right)=0 \qquad \tilde{J}_{ij} \,g\left({p_i^2}{ p_j^2}\right)=0 \qquad  J_{ij,kl}\,g(p_i^2 p_j^2 p_k^2 p_l^2)=0 \qquad J_{ij,kl}\, g\left( \frac{p_i^2}{ p_j^2}\frac{ p_k^2}{ p_l^2}\right)=0.
\label{lastref}
\end{equation}

Beside the exact solutions identified in the previous sections, the CWI's allows other classes of solutions which may be found using a limited set of assumptions on the $s,t$ dependence of the ans\"atz. Therefore, we will proceed with an analysis of the special CWI's, trying to find approximate solutions of Eqs. \eqref{C1}-\eqref{C3}. We will adopt the notations introduced in Eqs. \eqref{firstref}-\eqref{lastref} in order to refer to the various terms of the corresponding partial differential equations. For definiteness we will consider the case of Eq. \eqref{C1}, rewritten in the form \eqref{intform}. We will assume that $s$  and $t$ are both large invariants but we will keep their ratio fixed. In the Minkowski region this would correspond to investigating the contribution of such correlator for scatterings at fixed angle (i.e. $-t/s$ fixed).\\
We notice that if look for a factorised solution of the form 
\begin{equation}
\Phi(p_1,p_2,p_3,p_4)\equiv\chi(s,t) \phi(p_i^2),
\label{phi}
\end{equation}
 where we separate the dependence on the the external mass invariants $p_i^2$ from the $s,t$, we can satisfy the dilatation WI \eqref{dil1}  in the form 

\begin{align}
\label{dil21}
&\bigg[(\D_t-3d)-\sum_{i=1}^4p_i\frac{\partial}{\partial p_i}\bigg]\phi(p_i^2)=0\\
&\left(s\frac{\partial}{\partial s}+t\frac{\partial}{\partial t}\right)\chi(s,t)=0
\label{dil22}
\end{align}
with $\chi(s,t)\equiv \chi(s/t)$, i.e. an arbitrary function of the ratio of the two external invariants, describing energy and momentum transfers.\\
At this stage we can proceed with a separation of the special CWI \eqref{intform} into the three equations
\begin{align}
&D_{23}^{st}\chi(s/t)=0  \label{x1}\\
&(D^s J_{12,34}+ h_{34,12}D^s)\chi(s/t) \phi(p_i^2)=0 \label{x2}\\
&\textup{K}_{14}\phi=0 \label{x3}
\end{align}
of which we try to identify an asymptotic solution.\\
 Notice that a simple but exact solution of the first of the three equations above is logarithmic with $\chi(s/t)\sim \log(-t/s)$. It is also easy to check, by plugging this expression into the second equation, that 
 \begin{equation}
 \left( D^s J_{12,34}+ h_{34,12}D^s\right)\chi(s/t) \phi(p_i^2)\sim O(1/s^2,1/t^2) 
\end{equation}
and contributes insignificantly if the mass invariants $p_i^2$ stays bound. Indeed we will consider solutions of the ratios 
$p_i^2/p_j^2$ where this occurs. For this reason the solution of the last equation \eqref{x3} has to satisfy also \eqref{dil21}. We are clearly choosing to assign all the scaling behaviour of the global solution \eqref{phi} on the external mass invariants. If we require that  $p_i^2\sim O(1) \ll s, t$ then we can independently search for exact solutions of \eqref{x3}. 
\subsection{Factorized solutions as generalized hypergeometrics}
We can generalize these considerations to all the three CWI's \eqref{C1}, \eqref{C2}, \eqref{C3}, generating the system of equations 
\begin{equation}
\textup{K}_{14}\phi=0,\qquad \textup{K}_{24}\phi=0,\qquad \textup{K}_{34}\phi=0\label{CWILaur}
\end{equation}
where 
\begin{align}
\textup{K}_i&=\frac{\partial^2}{\partial p_i^2}+\frac{(d-2\D_i+1)}{p_i}\frac{\partial}{\partial p_i},\qquad i=1,\dots,4\ ,\\
\textup{K}_{ij}&=\textup{K}_i-\textup{K}_j\ .
\end{align}

An equivalent way to rearrange this operator is to use a change of variables from $(p_1^2,p_2^2,p_3^2,p_4^2)$ to $(x,y,z,p_4^2)$ where 
\begin{equation}
x=\sdfrac{p_1^2}{p_4^2},\quad y=\sdfrac{p_2^2}{p_4^2},\quad z=\sdfrac{p_3^2}{p_4^2}
\end{equation}
are the dimensionless rations x, y and z which must not to be confused with coordinate points in a three dimensional space. The ans\"atz for the solution can be taken of the form
\begin{equation}
\phi(p_1,p_2,p_3,p_4)=(p_4^2)^{n_s}\,x^a\,y^b\,z^c\,F(x,y,z),
\end{equation}
satisfying the dilatation Ward identity \eqref{dil21}

with the condition
\begin{equation}
n_s=\frac{\D_t}{2}-\frac{3d}{2}
\end{equation}
With this ans\"atz the conformal Ward identities read as
\begin{align}
\textup{K}_{14}\phi=&4p_4^{\D_t-3d-2}\,x^a\,y^b\,z^c\,\bigg[(1-x)x\sdfrac{\partial^2}{\partial x^2}-2x\,y\sdfrac{\partial^2}{\partial x\partial y}-y^2\sdfrac{\partial^2}{\partial y^2}-2x\,z\sdfrac{\partial^2}{\partial x\partial z}-z^2\sdfrac{\partial^2}{\partial z^2}-2y\,z\sdfrac{\partial^2}{\partial y\partial z}\notag\\
&\hspace{2cm}+(Ax+\gamma)\sdfrac{\partial}{\partial x}+Ay\sdfrac{\partial}{\partial y}+Az\sdfrac{\partial }{\partial z}+\left(E+\sdfrac{G}{x}\right)\bigg]F(x,y,z)=0
\end{align}
with
\begin{subequations}
	\begin{align}
	A&=\D_1+\D_2+\D_3-\sdfrac{5}{2}d-2(a+b+c)-1\\
	E&=-\sdfrac{1}{4}\big(3d-\D_t+2(a+b+c)\big)\big(2d+2\D_4-\D_t+2(a+b+c)\big)\\
	G&=\sdfrac{a}{2}\,\left(d-2\D_1+2a\right)\\
	\g&=\sdfrac{d}{2}-\D_1+2a+1
	\end{align}
\end{subequations}
Similar constraints are obtained from the equation $\textup{K}_{34}\phi=0$ that can be written as
\begin{align}
\textup{K}_{24}\phi=&4p_4^{\D_t-3d-2}\,x^a\,y^b\,z^c\,\bigg[-x^2\sdfrac{\partial^2}{\partial x^2}-2x\,y\sdfrac{\partial^2}{\partial x\partial y}+(1-y)y\sdfrac{\partial^2}{\partial y^2}-2x\,z\sdfrac{\partial^2}{\partial x\partial z}-z^2\sdfrac{\partial^2}{\partial z^2}-2y\,z\sdfrac{\partial^2}{\partial y\partial z}\notag\\
&\hspace{2cm}+A'x\sdfrac{\partial}{\partial x}+(A'y+\g')\sdfrac{\partial}{\partial y}+A'z\sdfrac{\partial }{\partial z}+\left(E'+\sdfrac{G'}{x}\right)\bigg]F(x,y,z)=0
\end{align}
with
\begin{subequations}
	\begin{align}
	A'&=\D_1+\D_2+\D_3-\sdfrac{5}{2}d-2(a+b+c)-1\\
	E'&=-\sdfrac{1}{4}\big(3d-\D_t+2(a+b+c)\big)\big(2d+2\D_4-\D_t+2(a+b+c)\big)\\
	G'&=\sdfrac{b}{2}\,\left(d-2\D_2+2b\right)\\
	\g'&=\sdfrac{d}{2}-\D_2+2b+1
	\end{align}
\end{subequations}
and finally, for the third condition coming from the conformal Ward identities
\begin{align}
\textup{K}_{34}\phi=&4p_4^{\D_t-3d-2}\,x^a\,y^b\,z^c\,\bigg[-x^2\sdfrac{\partial^2}{\partial x^2}-2x\,y\sdfrac{\partial^2}{\partial x\partial y}-y^2\sdfrac{\partial^2}{\partial y^2}-2x\,z\sdfrac{\partial^2}{\partial x\partial z}+(1-z)z\sdfrac{\partial^2}{\partial z^2}-2y\,z\sdfrac{\partial^2}{\partial y\partial z}\notag\\
&\hspace{2cm}+A''x\sdfrac{\partial}{\partial x}+A''y\sdfrac{\partial}{\partial y}+(A''z+\g'')\sdfrac{\partial }{\partial z}+\left(E''+\sdfrac{G''}{x}\right)\bigg]F(x,y,z)=0
\end{align}
with
\begin{subequations}
	\begin{align}
	A''&=\D_1+\D_2+\D_3-\sdfrac{5}{2}d-2(a+b+c)-1\\
	E''&=-\sdfrac{1}{4}\big(3d-\D_t+2(a+b+c)\big)\big(2d+2\D_4-\D_t+2(a+b+c)\big)\\
	G''&=\sdfrac{c}{2}\,\left(d-2\D_3+2c\right)\\
	\g''&=\sdfrac{d}{2}-\D_3+2c+1
	\end{align}
\end{subequations}

It is worth noticing that in order to perform the reduction to the hypergeometric form of the equations, we need to set $G=0$, $G'=0$ and $G''=0$, which imply that the Fuchsian points $a,b,c$ have different values as
\begin{subequations}
	\begin{align}
	a&=0,\,\D_1-\sdfrac{d}{2}\\
	b&=0,\,\D_2-\sdfrac{d}{2}\\
	c&=0,\,\D_3-\sdfrac{d}{2}.
	\end{align}
\end{subequations}
We find also that $E=E'=E''=-\a(a,b,c)\,\b(a,b,c)$ where
\begin{align}
\a(a,b,c)&=d+\D_4-\sdfrac{\D_t}{2}+a+b+c\notag\\
\b(a,b,c)&=\sdfrac{3d}{2}-\sdfrac{\D_t}{2}+a+b+c
\end{align}
as well as $A=A'=A''=-(\a(a,b,c)+\b(a,b,c)+1)$, indeed
\begin{align}
A=A'=A''&=-(\a(a,b,c)+\b(a,b,c)+1)=\D_1+\D_2+\D_3-\sdfrac{5}{2}d-2(a+b+c)-1
\end{align}
and finally
\begin{equation}
\g(a)=\frac{d}{2}-\Delta_1+2a+1\,,\qquad\g'(b)=\frac{d}{2}-\Delta_2+2b+1\,,\qquad\g''(c)=\frac{d}{2}-\Delta_3+2c+1.
\end{equation}
With this redefinition of the coefficients, the equations are then expressed in the form
\begin{equation}
\resizebox{1\hsize}{!}{$
\left\{
\begin{matrix}
&x_j(1-x_j)\sdfrac{\partial^2F}{\partial x_j^2}+\hspace{-1cm}\sum\limits_{\substack{\hspace{1.3cm}s\ne j\ \text{for}\ r=j}}\hspace{-1.1cm}x_r\hspace{0.2cm}\sum x_s\hspace{0.5ex}\sdfrac{\partial^2F}{\partial x_r\partial x_s}+\left[\g_j-(\a+\b+1)x_j\right]\sdfrac{\partial F}{\partial x_j}-(\a+\b+1)\sum\limits_{k\ne j}\,x_k\sdfrac{\partial F}{\partial x_k}-\a\,\b\,F=0\\[3ex]
& (j=1,2,3)
\end{matrix}\right.\label{systemLauricella}$}
\end{equation}
where for sake of simplicity we have re-defined $\g_1=\g$, $\ \g_2=\g'$ and $\g_3=\g''$ and $x_1=x$, $x_2=y$ and $x_3=z$. 
This system of equations allows solutions in the form of the Lauricella hypergeometric function $F_C$ of three variables, defined by the series 
\begin{equation}
F_C(\a,\b,\g,\g',\g'',x,y,z)=\sum\limits_{m_1,m_2,m_3}^\infty\,\frac{(\a)_{m_1+m_2+m_3}(\b)_{m_1+m_2+m_3}}{(\g)_{m_1}(\g')_{m_2}(\g'')_{m_3}m_1!\,m_2!\,m_3!}x^{m_1}y^{m_2}z^{m_3}.
\end{equation}
where the Pochhammer symbol $(\l)_{k}$ with an arbitrary $\l$ and $k$ a positive integer not equal to zero, was previously defined in \eqref{Pochh}. The convergence region of this series is defined by the condition
\begin{equation}
\left|\sqrt{x}\right|+\left|\sqrt{y}\right|+\left|\sqrt{z}\right|<1.
\end{equation}
The function $F_C$ is the generalization of the Appell $F_4$ for the case of three variables.
The system of equations \eqref{systemLauricella} admits 8 independent particular integrals (solutions) listed below
	\begin{align}
	&S_1(\a,\b,\g,\g',\g'',x,y,z)=F_C\big(\a,\b,\g,\g',\g'',x,y,z\big)\notag,\\
	&S_2(\a,\b,\g,\g',\g'',x,y,z)=x^{1-\g}\,F_C\big(\a-\g+1,\b-\g+1,2-\g,\g',\g'',x,y,z\big)\notag\,,\\
	&S_3(\a,\b,\g,\g',\g'',x,y,z)= y^{1-\g'}\,F_C\big(\a-\g'+1,\b-\g'+1,\g,2-\g',\g'',x,y,z\big)\notag\,,\\
	&S_4(\a,\b,\g,\g',\g'',x,y,z)=z^{1-\g''}\,F_C\big(\a-\g''+1,\b-\g''+1,\g,\g',2-\g'',x,y,z\big)\notag,\,\\
	&S_5(\a,\b,\g,\g',\g'',x,y,z)=x^{1-\g}y^{1-\g'}\,F_C\big(\a-\g-\g'+2,\b-\g-\g'+2,2-\g,2-\g',\g'',x,y,z\big)\,,\notag\\
	&S_6(\a,\b,\g,\g',\g'',x,y,z)=x^{1-\g}z^{1-\g''}\,F_C\big(\a-\g-\g''+2,\b-\g-\g''+2,2-\g,\g',2-\g'',x,y,z\big)\notag\,,\\
	&S_7(\a,\b,\g,\g',\g'',x,y,z)=y^{1-\g'}z^{1-\g''}\,F_C\big(\a-\g'-\g''+2,\b-\g'-\g''+2,\g,2-\g',2-\g'',x,y,z\big)\notag\,,\\
	&S_8(\a,\b,\g,\g',\g'',x,y,z)=x^{1-\g}y^{1-\g'}z^{1-\g''}\notag\\
	&\hspace{4cm}\times\,F_C\big(\a-\g-\g'-\g''+2,\b-\g-\g'-\g''+2,2-\g,2-\g',2-\g'',x,y,z\big)\,.\label{oneeq}
	\end{align} 
	where we have defined 
	\begin{align}
	\a&\equiv\a(0,0,0)=d+\D_4-\sdfrac{\D_t}{2}\notag\\	
	\b&\equiv\b(0,0,0)=\sdfrac{3d}{2}-\sdfrac{\D_t}{2}\notag\\
	\g&\equiv\g(0)=\frac{d}{2}-\Delta_1+1\notag\\
	\g'&\equiv\g'(0)=\frac{d}{2}-\Delta_2+1\notag\\
	\g''&\equiv\g''(0)=\frac{d}{2}-\Delta_3+1.
	\end{align}
Finally the solution for $\phi$  can be written as
\begin{equation}
\phi(p_i^2)=p_4^{\D_t-3d}\sum_i C_i\ S_i(\a,\b,\g,\g',\g'',x,y,z)
\end{equation}
where $C_i$ are arbitrary constant and $S_i$, $i=1,\dots, 2^3$ are the independent solutions written above. \\
To summarize, we have indeed shown that approximate solutions of the CWI's, describing the behaviour of the correlator at fixed angle 
can be taken of the factorized form 
\begin{equation}
\Phi(p_1,p_2,p_3,p_4)\sim \log(-t/s) \phi(p_i^2).
\end{equation}
We should remark that other approximate solutions of similar form, containing higher powers of logarithms of $-t/s$ are also compatible with the asymptotic ans\"atz that we have presented here. Obviously, in such a case we would be requiring that the exact condition \eqref{x1} would be replaced by the new condition 
\begin{equation}
D_{23}^{st}\chi(s/t)=O(1/s^2,1/t^2)  
\end{equation}
which is asymptotically satisfied also by higher powers of $\log(-t/s)$. In general, under such weaker assumptions, approximate asymptotic solutions can be summarized in the more general form
\begin{equation}
\Phi(p_1,p_2,p_3,p_4)\sim f\left(\log(-t/s) \right)\phi(p_i^2).
\end{equation}
where $f$ can be take of the generic form 
\begin{equation} 
f\left(\log(-t/s)\right) = \sum_k c_k \log^k\left((-t/s)\right).
\end{equation}
In the next section we are going to show that for the $p_i^2$ dependence on the external mass invariants of the approximate solution, given by the Lauricella functions, their equivalence to 4-K integrals, generalizing previous results for 3-point functions. 
\subsection{Lauricella's as 4-K integrals}
It is interesting to show how the solutions found above can be reformulated in a way which resembles what found in the case of 3-point functions. As alredy mentioned, the 3K integrals provide an efficient alternative way to express the solutions for scalar 3-point functions in terms of Appell functions. We are now going to show that  hypergeometrics of 3-variables, which belong to the class of Lauricella functions, similarly, can be related to 4K integrals. 
We write the solutions of such systems in the form
\begin{align}
I_{\a-1\{\n_1,\n_2,\n_3,\n_4\}}(a_1,a_2,a_3,a_4)&=\int_0^\infty\,dx\,x^{\a-1}\,\prod_{i=1}^4(a_i)^{\n_i}\,K_{\n_i}(a_i\,x)
\label{4Kintegral}
\end{align} 
with the Bessel functions $I_\nu,J_\nu, K_\nu$ related by the identities
\begin{align}
I_\nu(x)&=i^{-\n}\,J_{\n}(i\,x)\\
K_\nu(x)&=\frac{\pi}{2\sin(\pi\,\n)}\bigg[I_{-\n}(x)-I_\n(x)\bigg]=\frac{1}{2}\bigg[i^\nu\, \G(\n)\G(1-\n)\,J_{-\n}(i\,x)+i^{-\n}\,\G(-\n)\G(1+\n)\,J_\n(i\,x)\bigg]\label{Kscomp}
\end{align}
where we have used the properties of the Gamma functions
\begin{equation}
\frac{\pi}{\sin(\pi\n)}=\G(\n)\,\G(1-\n),\qquad-\frac{\pi}{\sin(\pi\n)}=\G(-\n)\,\G(1+\n).
\end{equation}

The structure of the CWI's \eqref{CWILaur} supports this formulation. The dilatation Ward identities in this case can be written as
\begin{equation}
\bigg[(\D_t-3d)-\sum_{i=1}^4p_i\frac{\partial}{\partial p_i}\bigg]I_{\a\{\b_1,\b_2,\b_3,\b_4\}}(p_1,p_2,p_3,p_4)=0
\end{equation}
and using the properties of 4K integrals in \appref{AppendixB} we derive the relation
\begin{align}
(\a-\b_t+1+\D_t-3d)I_{\a\{\b_1,\b_2,\b_3,\b_4\}}(p_1,p_2,p_3,p_4)=0
\end{align}
which is identically satisfied if the $\a$ exponent is equal to $\tilde{\a}$
\begin{equation}
\tilde{\a}=\b_t+3d-\D_t-1.
\end{equation}
The conformal Ward identities \eqref{CWILaur} can now be written as
\begin{equation}
\left\{
\begin{aligned}
\textup{K}_{14}I_{\tilde{\a}\{\b_1,\b_2,\b_3,\b_4\}}&=0\\
\textup{K}_{24}I_{\tilde{\a}\{\b_1,\b_2,\b_3,\b_4\}}&=0\\
\textup{K}_{34}I_{\tilde{\a}\{\b_1,\b_2,\b_3,\b_4\}}&=0,
\end{aligned}
\right.
\end{equation}
generating the final relations
\begin{equation}
\left\{
\begin{aligned}
(d-2\D_4+2\b_4)I_{\tilde{\a}+1\{\b_1,\b_2,\b_3,\b_4-1\}}-(d-2\D_1+2\b_1)I_{\tilde{\a}+1\{\b_1-1,\b_2,\b_3,\b_4\}}&=0\\
(d-2\D_4+2\b_4)I_{\tilde{\a}+1\{\b_1,\b_2,\b_3,\b_4-1\}}-(d-2\D_2+2\b_2)I_{\tilde{\a}+1\{\b_1,\b_2-1,\b_3,\b_4\}}&=0\\
(d-2\D_4+2\b_4)I_{\tilde{\a}+1\{\b_1,\b_2,\b_3,\b_4-1\}}-(d-2\D_3+2\b_3)I_{\tilde{\a}+1\{\b_1,\b_2,\b_3-1,\b_4\}}&=0\\
\end{aligned}
\right.
\end{equation}
which are satisfied if
\begin{align}
\b_i=\D_i-\frac{d}{2},\qquad i=1,\dots,4
\end{align}
giving
\begin{equation}
\tilde{\a}=d-1.
\end{equation}
The final solution can be written as
\begin{align}
\phi(p_1,p_2,p_3,p_4)&=\bar{\bar{\a}}\, I_{d-1\left\{\D_1-\frac{d}{2},\D_2-\frac{d}{2},\D_3-\frac{d}{2},\D_4-\frac{d}{2}\right\}}(p_1,p_2,p_3,p_4)\notag\\
&=\int_0^\infty\,dx\,x^{d-1}\,\prod_{i=1}^4(p_i)^{\D_i-\frac{d}{2}}\,K_{\D_i-\frac{d}{2}}(p_i\,x).
,\label{4Kfin}
\end{align}
where $\bar{\bar{\a}}$ is a undetermined constant. \\
Concerning the convergence of the approximate 4K solutions found in the fixed angle scattering limit at large $s$ and $t$, one can discuss the general conditions to be 
imposed, by following a strategy quite similar to the one discussed in \secref{convergence}.

The asymptotic limit at large and small x values, also in this case previously shown in 
\eqref{questoK} et seq., gives the conditions
\begin{align}
&p_1+p_2+p_3+p_4>0
\end{align}
and
\begin{align}
&\frac{d}{2}-1\pm\left(\Delta_1-\frac{d}{2}\right)\pm\left(\Delta_2-\frac{d}{2}\right)\pm\left(\Delta_3-\frac{d}{2}\right)\pm\left(\Delta_4-\frac{d}{2}\right)>0,
\end{align}
respectively. Therefore the condition of convergence at large $x$ of the parametric representation of the 4K integral is verified within the physical region of the general scalar 4-point function. Also in this case, as in \secref{convergence}, a discussion of implications of such convergence constraints will be presented in a related work.

\subsection{Connection with the Lauricella}
The key identity necessary to obtain the relation between the Lauricella functions and the 4K integral takes the form
\begin{align}
\int_0^\infty dx\,x^{\a-1}\prod_{j=1}^3\,J_{\m_j}(a_j\,x)\,K_{\nu}(c\,x)&=2^{\a-2}\,c^{-\a-\l}\,\G\left(\frac{\a+\l-\n}{2}\right)\G\left(\frac{\a+\l+\n}{2}\right)\notag\\
&\hspace{-3cm}\times\prod_{j=1}^3\,\frac{a_j^{\m_j}}{\G(\m_j+1)}F_C\left(\frac{\a+\l-\n}{2},\frac{\a+\l+\n}{2},\m_1+1,\m_2+1,\m_3+1;-\frac{a_1^2}{c^2},-\frac{a_2^2}{c^2},-\frac{a_3^2}{c^2}\right)\notag\\
&\centering\,\bigg[\l=\sum_{j=1}^3\,\m_j\,;\, \Re(\a+\l)>|\Re(\n)|,\,\Re(c)>\sum_{j=1}^{3}|\Im\,a_j|\bigg]\label{Prudnikov}.
\end{align}
One of the advantages of the use of the 4K integral expression of a solution is the simplified way by which the symmetry conditions can be imposed. In fact, by taking each of the 8 independent solutions identified in \eqref{oneeq}, and by rewriting them in the form of 4K integrals, we can impose the symmetry constraints far more easily.
 Then the general 4K integral in \eqref{4Kintegral}, using \eqref{Kscomp}, can be written as

\begin{equation}
\resizebox{1\hsize}{!}{$\begin{aligned}
\Phi(p_i^2)&=2^{d-5}p_4^{\D_t-3d}\, C_{1234}\bigg\{\,\G\left(\D_1-\frac{d}{2}\right)\G\left(\D_2-\frac{d}{2}\right)\G\left(\D_3-\frac{d}{2}\right)\,\G\left(\frac{3d-\D_t}{2}\right)\G\left(d+\D_4-\frac{\D_t}{2}\right)\\
&\hspace{1cm}\times F_C^{(3)}\left(\frac{3d}{2}-\frac{\D_t}{2}\,,\,d+\D_4-\frac{\D_t}{2},\frac{d}{2}-\D_1+1,\frac{d}{2}-\D_2+1,\frac{d}{2}-\D_3+1;\frac{p_1^2}{p_4^2},\frac{p_2^2}{p_4^2},\frac{p_3^2}{p_4^2}\right)\\[1.5ex]
&+\G\left(\D_1-\frac{d}{2}\right)\G\left(\D_2-\frac{d}{2}\right)\G\left(\frac{d}{2}-\D_3\right)\,\G\left(\frac{d}{2}-\frac{\D_t}{2}+\D_3+\D_4\right)\G\left(d+\D_3-\frac{\D_t}{2}\right)\left(\frac{p_3^2}{p_4^2}\right)^{\D_3-\frac{d}{2}}\\
&\hspace{1cm}\times F_C^{(3)}\left(d-\frac{\D_t}{2}+\D_3\,,\,\frac{d}{2}-\frac{\D_t}{2}+\D_3+\D_4,\frac{d}{2}-\D_1+1,\frac{d}{2}-\D_2+1,1-\frac{d}{2}+\D_3;\frac{p_1^2}{p_4^2},\frac{p_2^2}{p_4^2},\frac{p_3^2}{p_4^2}\right)\\[1.5ex]
&+\G\left(\D_1-\frac{d}{2}\right)\G\left(\frac{d}{2}-\D_2\right)\G\left(\D_3-\frac{d}{2}\right)\,\G\left(d-\frac{\D_t}{2}+\D_2\right)\G\left(\frac{d}{2}-\frac{\D_t}{2}+\D_2+\D_4\right)\left(\frac{p_2^2}{p_4^2}\right)^{\D_2-\frac{d}{2}}\\
&\hspace{1cm}\times F_C^{(3)}\left(d-\frac{\D_t}{2}+\D_2\,,\,\frac{d}{2}-\frac{\D_t}{2}+\D_2+\D_4,\frac{d}{2}-\D_1+1,1-\frac{d}{2}+\D_2,\frac{d}{2}-\D_3+1;\frac{p_1^2}{p_4^2},\frac{p_2^2}{p_4^2},\frac{p_3^2}{p_4^2}\right)\\[1.5ex]
&+\G\left(\frac{d}{2}-\D_1\right)\G\left(\D_2-\frac{d}{2}\right)\G\left(\D_3-\frac{d}{2}\right)\,\G\left(d-\frac{\D_t}{2}+\D_1\right)\G\left(\frac{d}{2}-\frac{\D_t}{2}+\D_1+\D_4\right)\left(\frac{p_1^2}{p_4^2}\right)^{\D_1-\frac{d}{2}}\\
&\hspace{1.5cm}\times F_C^{(3)}\left(d-\frac{\D_t}{2}+\D_1\,,\,\frac{d}{2}-\frac{\D_t}{2}+\D_1+\D_4,1-\frac{d}{2}+\D_1,\frac{d}{2}-\D_2+1,\frac{d}{2}-\D_3+1;\frac{p_1^2}{p_4^2},\frac{p_2^2}{p_4^2},\frac{p_3^2}{p_4^2}\right)\\[1.5ex]
&+\G\left(\D_1-\frac{d}{2}\right)\G\left(\frac{d}{2}-\D_2\right)\G\left(\frac{d}{2}-\D_3\right)\,\G\left(\frac{\D_t}{2}-\D_1\right)\G\left(\frac{d}{2}-\frac{\D_t}{2}+\D_2+\D_3\right)\left(\frac{p_2^2}{p_4^2}\right)^{\D_2-\frac{d}{2}}\left(\frac{p_3^2}{p_4^2}\right)^{\D_3-\frac{d}{2}}\\
&\hspace{1.5cm}\times F_C^{(3)}\left(\frac{d}{2}-\frac{\D_t}{2}+\D_2+\D_3\,,\,\frac{\D_t}{2}-\D_1,1+\frac{d}{2}-\D_1,1-\frac{d}{2}+\D_2,1-\frac{d}{2}+\D_3;\frac{p_1^2}{p_4^2},\frac{p_2^2}{p_4^2},\frac{p_3^2}{p_4^2}\right)\notag\\
\end{aligned}$}
\end{equation}
\begin{equation}
\resizebox{1\hsize}{!}{$
\begin{aligned}
&+\G\left(\frac{d}{2}-\D_1\right)\G\left(\frac{d}{2}-\D_2\right)\G\left(\D_3-\frac{d}{2}\right)\,\G\left(\frac{\D_t}{2}-\D_3\right)\G\left(\frac{d}{2}-\frac{\D_t}{2}+\D_1+\D_2\right)\left(\frac{p_1^2}{p_4^2}\right)^{\D_1-\frac{d}{2}}\left(\frac{p_2^2}{p_4^2}\right)^{\D_2-\frac{d}{2}}\\[1.5ex]
&\hspace{1.5cm}\times F_C^{(3)}\left(\frac{\D_t}{2}-\D_3\,,\,\frac{d}{2}-\frac{\D_t}{2}+\D_1+\D_2,1+\D_1-\frac{d}{2},1+\frac{d}{2}-\D_2,1-\frac{d}{2}+\D_3;\frac{p_1^2}{p_4^2},\frac{p_2^2}{p_4^2},\frac{p_3^2}{p_4^2}\right)\\[1.5ex]
&+\G\left(\frac{d}{2}-\D_1\right)\G\left(\D_2-\frac{d}{2}\right)\G\left(\frac{d}{2}-\D_3\right)\,\G\left(\frac{\D_t}{2}-\D_2\right)\G\left(\frac{d}{2}-\frac{\D_t}{2}+\D_1+\D_3\right)\left(\frac{p_1^2}{p_4^2}\right)^{\D_1-\frac{d}{2}}\left(\frac{p_3^2}{p_4^2}\right)^{\D_3-\frac{d}{2}}\\
&\hspace{1.5cm}\times F_C^{(3)}\left(\frac{\D_t}{2}-\D_2\,,\,\frac{d}{2}-\frac{\D_t}{2}+\D_1+\D_3,1+\D_1-\frac{d}{2},1+\frac{d}{2}-\D_2,1-\frac{d}{2}+\D_3;\frac{p_1^2}{p_4^2},\frac{p_2^2}{p_4^2},\frac{p_3^2}{p_4^2}\right)\\
&+\G\left(\frac{d}{2}-\D_1\right)\G\left(\frac{d}{2}-\D_2\right)\G\left(\frac{d}{2}-\D_3\right)\,\G\left(\frac{3d}{2}-\frac{\D_t}{2}\right)\G\left(d-\frac{\D_t}{2}+\D_4\right)\left(\frac{p_1^2}{p_4^2}\right)^{\D_1-\frac{d}{2}}\left(\frac{p_2^2}{p_4^2}\right)^{\D_2-\frac{d}{2}}\left(\frac{p_3^2}{p_4^2}\right)^{\D_3-\frac{d}{2}}\\
&\hspace{0.5cm}\times F_C^{(3)}\left(\frac{3d}{2}-\frac{\D_t}{2}\,,\,d-\frac{\D_t}{2}+\D_4,1+\D_1-\frac{d}{2},1+\D_2-\frac{d}{2},1-\frac{d}{2}+\D_3;\frac{p_1^2}{p_4^2},\frac{p_2^2}{p_4^2},\frac{p_3^2}{p_4^2}\right)\bigg\}
\end{aligned}$}
\end{equation}
where $C_{1234}$ is the only undetermined constant.

\section{Conclusions}
We have investigated two classes of solutions of the CWI's of scalar primary correlators in momentum space. In the first class we have identified solutions in the form of 4-point functions which are dual conformal and conformal at the same time. Such solutions have been found using ans\"atze which allow to reduce the equations to systems of generalized hypergeometrics. The method extends previous analysis of 3-point functions for scalar and tensor correlators, limitedly to DCC solutions, which can be expressed in terms of 3K integrals, similarly to the case of ordinary 3-point functions. \\ 
We have also discussed how one can construct solutions of the CWI's, by showing that at large $s$ and $t$, with a fixed $-t/s$, i.e. at fixed angle, the CWI's are approximated by a system of special hypergeometric equations, which can be solved by a specific factorized ansatz. In the ansatz, which is an exact solution of such a system, the dependence of the correlators on the external mass invariants is separated from the the $s$ and $t$ invariants. We have shown that  the solutions, in this case, take the form of Lauricella hypergeometric functions of 3 variables. The $s$ and $t$ dependence of the solutions is compatible with the structure of such correlators at fixed angle angle in the asymptotic limit, due to the logarithmic $-t/s$ dependence, typical of such solutions.\\
Finally, we have shown that the system of the Lauricella solutions are equivalent to some newly introduced 4K integrals. Would be very interesting to investigate whether this pattern can be extended to n-point functions, in the context of more realistic field theories such as QCD, for instance, following the analysis presented in \cite{Kidonakis:1998nf,Sterman:2002qn,Aybat:2006mz}.

  \centerline{\bf Acknowledgments} 
 We thank George Sterman for discussions and comments. This work is partly supported by INFN under Iniziative Specifica QFT-HEP. 

\appendix

\section{Chain rules}\label{Appendix0}
In this section we summarize some important relations regarding the chain rules used in the derivation of the hypergeometric system of equations. They are given by
\begin{align}
\frac{\partial^2}{\partial p_1^2} F(x,y)&=\frac{2x}{p_1^2}\ \partial_x F(x,y)+\frac{4x^2}{p_1^2}\ \partial_{xx}F,&& \frac{\partial^2}{\partial p_4^2} F(x,y)=\frac{2y}{p_4^2}\ \partial_y F(x,y)+\frac{4y^2}{p_4^2}\ \partial_{yy}F,\\[1.3ex]
\frac{\partial^2}{\partial p_3^2} F(x,y)&=\frac{2x}{p_3^2}\ \partial_x F(x,y)+\frac{4x^2}{p_3^2}\ \partial_{xx}F,&& \frac{\partial^2}{\partial p_2^2} F(x,y)=\frac{2y}{p_2^2}\ \partial_y F(x,y)+\frac{4y^2}{p_2^2}\ \partial_{yy}F,\\[1.3ex]
\frac{\partial}{\partial p_1} F(x,y)&=\frac{2x}{p_1} \ \partial_x F(x,y),&&\frac{\partial}{\partial p_4} F(x,y)=\frac{2y}{p_4} \ \partial_y F(x,y),\\[1.3ex]
\frac{\partial}{\partial p_3} F(x,y)&=\frac{2x}{p_3} \ \partial_x F(x,y),&&\frac{\partial}{\partial p_2} F(x,y)=\frac{2y}{p_2} \ \partial_y F(x,y),\\[1.3ex]
\frac{\partial}{\partial s}F(x,y)&=-\frac{2}{s}\big(x\,\partial_x F+y\,\partial_yF\big),&&\frac{\partial}{\partial t}F(x,y)=-\frac{2}{t}\big(x\,\partial_x F+y\,\partial_yF\big),
\end{align}
\begin{equation}
\frac{\partial^2}{\partial s\partial t}F(x,y)=\frac{4}{st}\big[\big(
x\,\partial_x +y\partial_y\big)F+\big(x^2\partial_{xx}+2xy\,\partial_{xy}+y^2\partial_{yy}\big)F\big],
\end{equation}
\begin{align}
\left(p_1\frac{\partial}{\partial p_1}+p_2\frac{\partial}{\partial p_2}-p_3\frac{\partial}{\partial p_3}-p_4\frac{\partial}{\partial p_4}\right)F(x,y)&=\left(2x\,\partial_x+2y\,\partial_y -2x\,\partial_x-2y\,\partial_y\right)F(x,y)=0,\\[1.5ex]
\left(p_1\frac{\partial}{\partial p_1}+p_4\frac{\partial}{\partial p_4}-p_3\frac{\partial}{\partial p_3}-p_2\frac{\partial}{\partial p_2}\right)F(x,y)&=\left(2x\,\partial_x+2y\,\partial_y -2x\,\partial_x-2y\,\partial_y\right)F(x,y)=0.
\end{align}

\section{3K integrals for 4-point functions}\label{AppendixA}
We summarize some relations concerning 3K integrals. We define 
\begin{equation}
I_{\a\{\b_1,\b_2,\b_3\}}(p_1\,p_3; p_2\,p_4;s\,t)=\int_0^\infty\,dx\,x^\a\,(p_1p_3)^{\b_1}\,(p_2p_4)^{\b_2}\,(s\,t)^{\b_3}\,K_{\b_1}(p_1p_3\,x)\,K_{\b_2}(p_2p_4\,x)\,K_{\b_3}(st\,x)
\end{equation}
as in \eqref{3K}. The $K$ Bessel functions satisfy the relations
\begin{align}
\frac{\partial}{\partial p}\big[p^\b\,K_\b(p\,x)\big]&=-x\,p^\b\,K_{\b-1}(p x)\\
K_{\b+1}(x)&=K_{\b-1}(x)+\frac{2\b}{x}K_{\b}(x)
\end{align}
from which we obtain (omitting the argument in each integral as in \eqref{3K}) 
\begin{align}
\frac{\partial}{\partial p_1}I_{\a\{\b_1,\b_2,\b_3\}}&=-p_1\,p_3^2\,I_{\a+1\{\b_1-1,\b_2,\b_3\}}\\
\frac{\partial}{\partial p_3}I_{\a\{\b_1,\b_2,\b_3\}}&=-p_3\,p_1^2\,I_{\a+1\{\b_1-1,\b_2,\b_3\}}\\
\frac{\partial}{\partial p_2}I_{\a\{\b_1,\b_2,\b_3\}}&=-p_2\,p_4^2\,I_{\a+1\{\b_1,\b_2-1,\b_3\}}\\
\frac{\partial}{\partial p_4}I_{\a\{\b_1,\b_2,\b_3\}}&=-p_4\,p_2^2\,I_{\a+1\{\b_1,\b_2-1,\b_3\}}\\
\frac{\partial}{\partial s}I_{\a\{\b_1,\b_2,\b_3\}}&=-s\,t^2\,I_{\a+1\{\b_1,\b_2,\b_3-1\}}\\
\frac{\partial}{\partial t}I_{\a\{\b_1,\b_2,\b_3\}}&=-t\,s^2\,I_{\a+1\{\b_1,\b_2,\b_3-1\}}
\end{align}
and for the second derivative
\begin{align}
\frac{\partial^2}{\partial p_1^2}I_{\a\{\b_1,\b_2,\b_3\}}&=-\,p_3^2\,I_{\a+1\{\b_1-1,\b_2,\b_3\}}+p_1^2\,p_3^4\,\,I_{\a+2\{\b_1-2,\b_2,\b_3\}}\\
\frac{\partial^2}{\partial p_3^2}I_{\a\{\b_1,\b_2,\b_3\}}&=-\,p_1^2\,I_{\a+1\{\b_1-1,\b_2,\b_3\}}+p_3^2\,p_1^4\,\,I_{\a+2\{\b_1-2,\b_2,\b_3\}}\\
\frac{\partial^2}{\partial p_2^2}I_{\a\{\b_1,\b_2,\b_3\}}&=-\,p_4^2\,I_{\a+1\{\b_1,\b_2-1,\b_3\}}+p_2^2\,p_4^4\,\,I_{\a+2\{\b_1,\b_2-2,\b_3\}}\\
\frac{\partial^2}{\partial p_4^2}I_{\a\{\b_1,\b_2,\b_3\}}&=-\,p_2^2\,I_{\a+1\{\b_1,\b_2-1,\b_3\}}+p_4^2\,p_2^4\,\,I_{\a+2\{\b_1,\b_2-2,\b_3\}}\\
\frac{\partial^2}{\partial s\partial t}I_{\a\{\b_1,\b_2,\b_3\}}&=-2\,s\,t\,I_{\a+1\{\b_1,\b_2,\b_3-1\}}+t^3\,s^3\,\,I_{\a+2\{\b_1,\b_2,\b_3-2\}}.
\end{align}
They can be rearranged using the relations
\begin{align}
p_1^2\,p_3^2\,I_{\a+2\{\b_1-2,\b_2,\b_3\}}&=I_{\a+2\{\b_1,\b_2,\b_3\}}-2(\b_1-1)\,I_{\a+1\{\b_1-1,\b_2,\b_3\}}\\
p_2^2\,p_4^2\,I_{\a+2\{\b_1,\b_2-2,\b_3\}}&=I_{\a+2\{\b_1,\b_2,\b_3\}}-2(\b_2-1)\,I_{\a+1\{\b_1,\b_2-1,\b_3\}}\\
s^2\,t^2\,I_{\a+2\{\b_1,\b_2,\b_3-2\}}&=I_{\a+2\{\b_1,\b_2,\b_3\}}-2(\b_3-1)\,I_{\a+1\{\b_1,\b_2,\b_3-1\}}.
\end{align}

\section{4K integrals for Lauricella 4-point functions}\label{AppendixB}
We summarize some important relations about the 4K integrals. Defining the 4K integral as
\begin{equation}
I_{\a\{\b_1,\b_2,\b_3,\b_4\}}(p_1,p_2,p_3,p_4)=\int_0^\infty\,dx\,x^\a\,\prod_{i=1}^4(p_i)^{\b_i}\,K_{\b_i}(p_i\,x)
\end{equation}
its first derivative with respect the mgnitudes of the momenta is given by
\begin{equation}
p_i\frac{\partial}{\partial p_i}I_{\a\{\b_j\}}=-p_i^2\,I_{\a+1\{\b_j-\d_{ij}\}},\qquad i,j=1,\dots,4.
\end{equation}
One can show that the relation
\begin{equation}
\int_0^\infty\,x^{\a+1}\frac{\partial}{\partial x}\left[\prod_{i=1}^4\,p_i^{\b_i}\,K_{\b_i}(p_i\,x)\right]=-\int_0^\infty\,\left[\frac{\partial x^{\a+1}}{\partial x}\right]\prod_{i=1}^4\,p_i^{\b_i}\,K_{\b_i}(p_i\,x)
\end{equation}
leads to the identity
\begin{equation}
\sum_{i=1}^{4}p_i^2I_{\a+1\{\b_j-\d_{ij}\}}=(\a-\b_t+1)\,I_{\a\{\b_j\}},\qquad j=1,\dots,4
\end{equation}
where $\b_t=\b_1+\b_2+\b_3+\b_4$.


\end{document}